\documentclass[acmsmall,screen,nonacm]{acmart}
\usepackage[inline]{enumitem}
\usepackage{amsmath}
\usepackage{placeins}
\usepackage{xspace}

\usepackage[normalem]{ulem} \usepackage{subcaption} \usepackage{wrapfig} \usepackage{tabularx} \usepackage{multirow} \usepackage{algorithm} \usepackage{algpseudocode}   
\newcounter{finding}
\newcounter{subfinding}[finding]
\renewcommand{\thesubfinding}{\thefinding(\alph{subfinding})}
\newcommand{\findingbox}[1]{{\par\vspace{3pt}\noindent\setlength{\fboxrule}{0.6pt}\fcolorbox{black}{gray!12}{\parbox{\dimexpr\linewidth-2\fboxsep-2\fboxrule\relax}{#1}}}\par\vspace{3pt}}
\newcommand{\finding}[2][]{%
  \findingbox{%
    \refstepcounter{finding}%
    \if\relax\detokenize{#1}\relax\else\label{#1}\fi
    \textbf{Finding \#\thefinding:} #2}}
\newcommand{\findingCB}[3][]{%
  \findingbox{%
    \refstepcounter{finding}%
    \if\relax\detokenize{#1}\relax\else\label{#1}\fi
    \textbf{Common Belief:} #2\par\smallskip
    \textbf{Finding \#\thefinding:} #3}}
\newcommand{\findingpair}[5][]{%
  \findingbox{%
    \refstepcounter{finding}%
    \if\relax\detokenize{#1}\relax\else\label{#1}\fi
    \refstepcounter{subfinding}\label{#2}%
    \textbf{Finding \#\thesubfinding:} #3\par\medskip
    \refstepcounter{subfinding}\label{#4}%
    \textbf{Finding \#\thesubfinding:} #5}}
\newcommand{\findingpairCB}[6][]{%
  \findingbox{%
    \refstepcounter{finding}%
    \if\relax\detokenize{#1}\relax\else\label{#1}\fi
    \textbf{Common Belief:} #2\par\smallskip
    \refstepcounter{subfinding}\label{#3}%
    \textbf{Finding \#\thesubfinding:} #4\par\medskip
    \refstepcounter{subfinding}\label{#5}%
    \textbf{Finding \#\thesubfinding:} #6}}

\newif\ifeditcolors
\editcolorsfalse
\newcommand{\edittext}[2]{{\ifeditcolors\color{#1}\fi{#2}}}

\newcommand{\huanedit}[1]{\edittext{purple}{#1}}

\newcommand{\eg}{{\it e.g.,}\xspace}
\newcommand{\ie}{{\it i.e.,}\xspace}

\newcommand{\rqone}{Which kernel-level factors govern NPU efficiency during vision encoding and prefill?}
\newcommand{\rqtwo}{Which kernel-level factors determine whether the CPU or NPU performs better during autoregressive decoding?}
\newcommand{\rqthree}{Which input resolutions yield substantial gains in task accuracy with modest TTFT overhead?}
\begin{document}

\title{How Far Are We From On-Device VLM Inference? A Measurement Study on Edge NPUs}

\author{Huan Yang}
\email{yanghuan9812@csu.edu.cn}
\affiliation{%
  \institution{Central South University}
  \city{Changsha}
  \country{China}
}
\author{Shiqi Jiang}
\email{shijiang@microsoft.com}
\affiliation{%
  \institution{Microsoft Research}
  \city{Shanghai}
  \country{China}
}
\author{Yu Mao}
\email{maoyu26@csu.edu.cn}
\affiliation{%
  \institution{Central South University}
  \city{Changsha}
  \country{China}
}
\author{Chuanyue Xiong}
\email{cyxiong@csu.edu.cn}
\affiliation{%
  \institution{Central South University}
  \city{Changsha}
  \country{China}
}
\author{Deyu Zhang}
\authornote{Corresponding author.}
\email{zdy876@csu.edu.cn}
\affiliation{%
  \institution{Central South University}
  \city{Changsha}
  \country{China}
}
\renewcommand{\shortauthors}{Yang et al.}

\begin{abstract}
Diverse optimizations spanning model architecture, inference runtime, and hardware acceleration have been proposed for on‑device vision‑language model (VLM) deployment on edge NPUs. Still, cross‑layer coordination of VLM execution on edge NPUs remains largely underexplored. This paper presents an empirical measurement study across six representative VLMs, three real‑world edge platforms, and ten multimodal benchmarks. Our evaluation drills down from end‑to‑end application‑level latency and thermal traces down to low‑level kernel‑scale execution behaviors, challenging several widely held community assumptions. First, softmax dominates attention costs; even lookup‑table‑based softmax optimizations cannot eliminate pipeline bubbles that severely underutilize NPU matrix units. Second, theoretical speedups from linear attention vanish under real‑NPU execution due to prohibitive tensor‑rearrangement overhead. Third, no hardware backend universally dominates autoregressive decoding: attention‑tensor‑shaped parallelism dictates NPU efficiency and shifts the CPU‑NPU performance balance. {We further find task-specific Sweet Resolutions for VLMs, which deliver meaningful accuracy gains at only modest latency overhead.} Complementing these findings, we provide actionable cross‑layer recommendations for hardware architects, runtime developers, and model designers. 
\end{abstract}

\maketitle


\section{Introduction}
\label{sec:introduction}

On-device inference for vision-language models (VLMs) is emerging as a dominant trend in the development of intelligent edge devices~\cite{qu2025_survey}. On one hand, human-agent interaction is rapidly evolving beyond text-only communication to incorporate visual and acoustic signals, enabling more natural and immersive multimodal interaction. On the other hand, running VLMs locally could reduce the privacy risks associated with uploading user data to cloud servers, as all such data are processed and stored on devices~\cite{yu2026_personalizedsurvey}. {Beyond that, on-device inference could also potentially provide a more responsive user experience by avoiding cloud communication delays.}



Driven by this trend, researchers have explored a wide range of approaches, including {new hardware designs to improve the computational capabilities of edge devices and optimizations to model architectures and inference runtimes to address resource constraints. These efforts aim to enable efficient on-device VLM inference.} For example, in hardware design, modern mobile NPUs are integrated into systems-on-chip (SoCs) and include dedicated accelerators optimized for matrix computations~\cite{hanpu}. These accelerators execute the computationally intensive general matrix multiplication (GEMM) operations in Transformers, offloading from general-purpose processors, \ie CPUs and mobile GPUs. In model design, new architectures, \eg those based on linear attention~\cite{lineartransformer}, have also been proposed. Linear attention reduces the quadratic computational cost of full attention and replaces its growing KV cache with a fixed-size recurrent state. In kernel implementation, tiled attention kernels such as FlashAttention improve memory access efficiency and on-chip cache reuse \cite{dao2022flashattentionfastmemoryefficientexact}.


Despite these considerable advances across areas, a critical open question remains: \textbf{how far are we from practical on-device VLM inference?} {Prior measurement studies~\cite{laskaridis2024meltingpointmobileevaluation, xiao2026understandinglargelanguagemodels} have sought to provide a first look at the end-to-end LLM inference performance on edge. However, few studies systematically examine models, inference runtimes, and hardware specifications in detail, analyze kernel-level performance, and identify performance gaps and their root causes. Such analyses are particularly scarce for vision-language models, which introduce vision-preprocessing overhead and multimodal execution patterns absent in text-only LLMs.}




{To this end, in this paper, we conduct thorough measurements over six popular VLM models on typical modern edge NPUs, as described in \S~\ref{sec:overall-performance}.} We observe that memory capacity may not be a barrier to VLM deployment on edge devices, as recent devices provide sufficient unified memory and this capacity could be readily expanded. However, end-to-end inference latency remains prohibitive for real-world deployment. In addition, sustained heavy computation during continuous inference causes substantial heat accumulation, triggering severe thermal throttling and degrading performance. {Based on our observations, we argue that all these limitations stem from a core bottleneck: poor coordination among model architecture designs, inference runtime optimizations, and hardware utilization.}

To systematically dissect the cross-layer efficiency issue, unlike prior work, we  adopt a kernel-centric perspective. {Kernels form a critical intermediate layer that connects high-level model architectures to low-level hardware implementations, where we could ultimately {reveal} the effects of poor cross-layer coordination. In this paper, {we} begin by investigating kernel-level bottlenecks in the prefill and decoding phases to formulate and answer our first two research questions (RQs):
}
\begin{itemize}
  \item \textbf{RQ1 (\S~\ref{sec:kernel-bottleneck}):} \rqone
  \item \textbf{RQ2 (\S~\ref{sec:decode-performance}):} \rqtwo
\end{itemize}


{Besides that, we also identify and investigate a VLM-specific performance lever for balancing task accuracy and inference latency:}
\begin{itemize}
  \item \textbf{RQ3 (\S~\ref{sec:resolution}):} \rqthree
\end{itemize}



To answer RQ1, we perform kernel-level profiling of vision encoding and prefill on edge NPUs to identify the main efficiency bottlenecks (\S~\ref{sec:kernel-bottleneck}). For RQ2, we compare kernel-level latency between CPU and NPU backends during autoregressive decoding and investigate how attention shapes and K/V context length affect their relative performance (\S~\ref{sec:decode-performance}). For RQ3, we characterize how input preprocessing strategies affect TTFT and identify task-specific \emph{sweet resolutions} to guide input resolution selection across multiple widely-used benchmarks (\S~\ref{sec:resolution}).



{Our kernel-driven profiling yields several counterintuitive empirical insights. We identify critical cross-layer bottlenecks that limit NPU-accelerated VLM performance and characterize \huanedit{how} the attention-related hardware parallelism affects the relative performance of CPU and NPU backends during autoregressive decoding. We also identify task-specific sweet resolutions that yield substantial accuracy gains with modest TTFT overhead, providing practical guidance on selecting input resolutions for VLM deployment on edge devices.}

We summarize our main contributions as follows:

\begin{itemize}
\item \textit{Kernel‑centric measurement methodology.} We build an end-to-end evaluation framework for edge-NPU VLM inference combining kernel profiling, latency/thermal characterization, and accuracy assessment, with systematic experiments over six VLMs, three platforms, and ten multimodal benchmarks.
\item \textit{Counter‑intuitive hardware profiling results.} Real-device measurements uncover cross-layer bottlenecks and revise prevailing assumptions: softmax, not GEMM, dominates attention overhead and causes NPU pipeline bubbles; tensor-layout overhead erases theoretical linear-attention gains; attention tensor parallelism controls CPU–NPU relative decode performance.
\item {\textit{Deployment insight: task-specific sweet resolutions.} We discover task-specific Sweet Resolutions for VLMs, improving accuracy under modest growth of latency overhead.}
\item {\textit{Actionable cross-layer suggestions.} We provide empirically grounded optimization recommendations for hardware, runtime, and VLM developers to support cross-layer co-design.}
\end{itemize}


The rest of this paper is organized as follows.
Section~\ref{sec:background} provides the necessary background.
Section~\ref{sec:methodology} details our evaluation setup and measurement protocols.
Section~\ref{sec:overall-performance} summarizes NPU inference throughput, memory footprints and thermal behaviors.
Section~\ref{sec:kernel-bottleneck} analyzes NPU bottlenecks during vision encoding and prefill.
Section~\ref{sec:decode-performance} examines performance differences between CPU and NPU backends during decoding.
Section~\ref{sec:resolution} investigates resolution-dependent latency–accuracy trade-offs and introduces Sweet Resolutions.
Section~\ref{sec:recommendations} presents cross-layer suggestions for hardware, runtime, and model developers.
Section~\ref{sec:related-work} reviews related work.
Section~\ref{sec:limitations} discusses the limitations of our study.
Finally, Section~\ref{sec:conclusion} concludes the paper.

\section{Background}
\label{sec:background}

VLM inference performance on edge NPUs is jointly governed by model architecture, the inference runtime and the hardware specifications. These three layers determine the computational workload, its implementation, and its mapping to specialized compute units, respectively. In this section, we provide  the background on each layer.

\subsection{Model Architectures of VLMs}
\label{sec:vlm-arch}

\subsubsection{Encoder-based and encoder-free VLMs.}

Most current VLMs adopt an encoder-based architecture, in which a dedicated vision encoder, \eg ViT, extracts visual features from input images, and a projection module aligns these features with the input representation space of the LLM backbone~\cite{mobilevlmv2,Vasu_2025_CVPR,lu2025bluelm}. This architecture leverages pre-trained visual representations and allows vision and language components to be selected independently~\cite{mobilevlmv2,diao2026pixels}. However, vision encoders often impose substantial computational and memory overhead, particularly on resource-constrained edge devices. For instance, the ViT-based vision encoder and projector in Qwen3.5-2B together contain over 330M parameters, as reported in Table~\ref{tab:models}. Representative encoder-based VLMs include Qwen3.5, Gemma4, SmolVLM2, and InternVL3.5~\cite{smolvlm,internvl35,qwen3.5,gemma4}.

Recently, several encoder-free VLMs have been proposed~\cite{diao2026pixels,11093135}. These models replace the standalone vision encoder with a lightweight embedding module and process visual and textual tokens within a shared multimodal backbone. {Integrating visual feature extraction into the backbone eliminates the separate vision-encoding stage and can reduce encoding overhead. Representative models include NEO1.5 and Mono-InternVL}~\cite{diao2026pixels,11093135}.


\subsubsection{Attention Mechanisms in VLMs.}
\label{sec:attn-arch}

Beyond the high-level architectural differences between encoder-based and encoder-free designs, VLMs employ distinct attention mechanisms, which are major determinants of inference performance. Two mechanisms are widely adopted: full attention and linear attention.


Full attention computes query–key scores over all positions permitted by the attention mask, applies softmax normalization, and aggregates the corresponding values; this procedure is known as scaled dot‑product attention (SDPA)~\cite{vaswani2017attentionisallyouneed}. FlashAttention~\cite{dao2022flashattentionfastmemoryefficientexact} optimizes SDPA by tiling and online softmax to reduce intermediate memory traffic without materializing the full attention matrix. VLMs, such as SmolVLM2 and InternVL3.5~\cite{smolvlm,internvl35}, adopt full attention mechanism.


Linear attention avoids the quadratic computational cost and growing KV cache of full attention by compressing past context into a fixed-size recurrent state~\cite{lineartransformer}. Gated DeltaNet (GDN) is one such variant~\cite{yang2025gdn}. Qwen3.5~\cite{qwen3.5}, for instance, combines GDN with full-attention layers to enable efficient long-context processing.

\subsection{Hardware on Edge Devices}

\subsubsection{Asynchronous Computation on NPUs.}
\begin{figure}[htbp]
  \centering
  \begin{subfigure}[b]{0.45\textwidth}
    \centering
    \includegraphics[width=\linewidth]{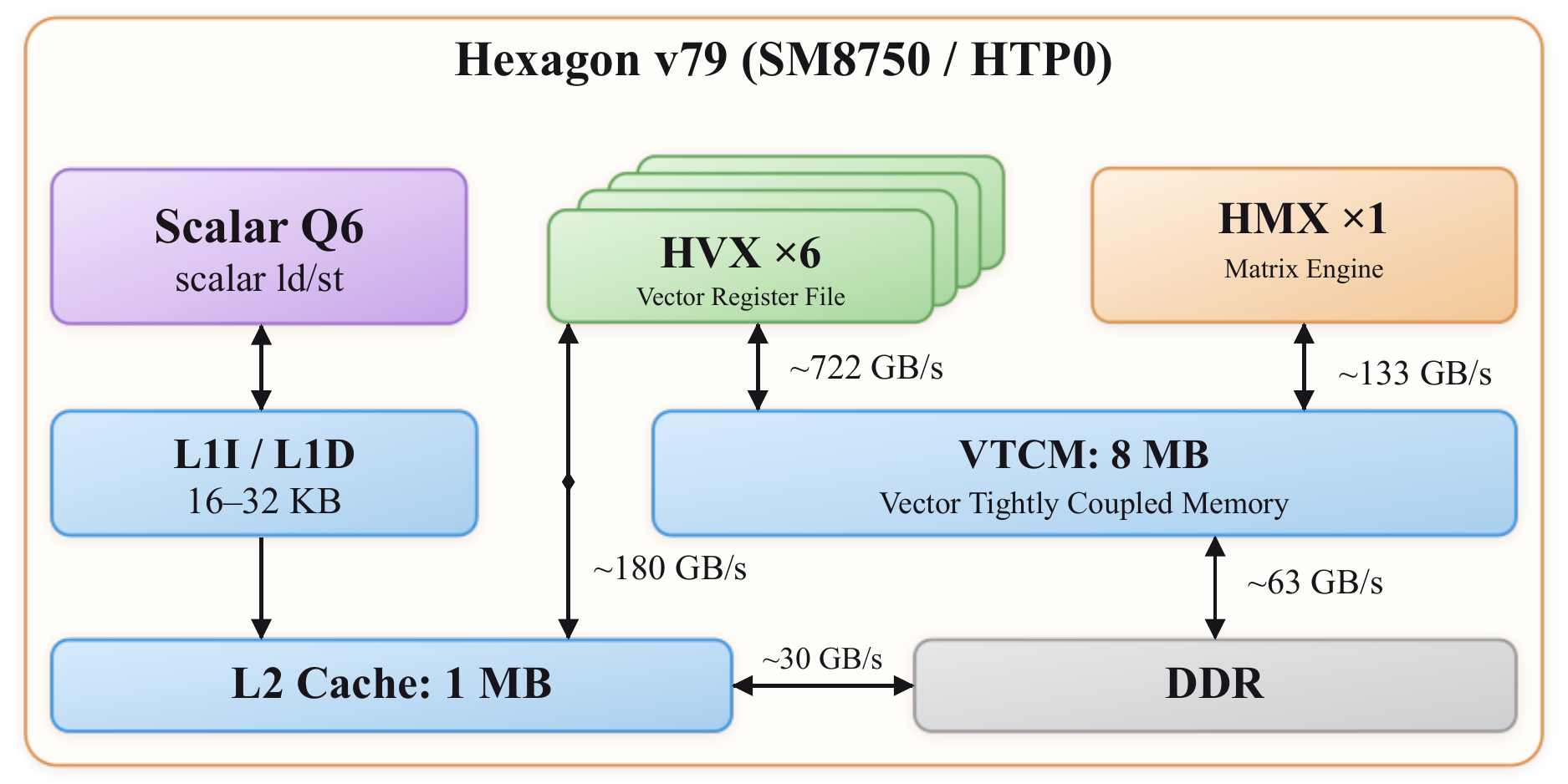}
    \caption{Qualcomm Hexagon v79 (SM8750).}
    \label{fig:backends-hx}
  \end{subfigure}\hfill
  \begin{subfigure}[b]{0.45\textwidth}
    \centering
    \includegraphics[width=\linewidth]{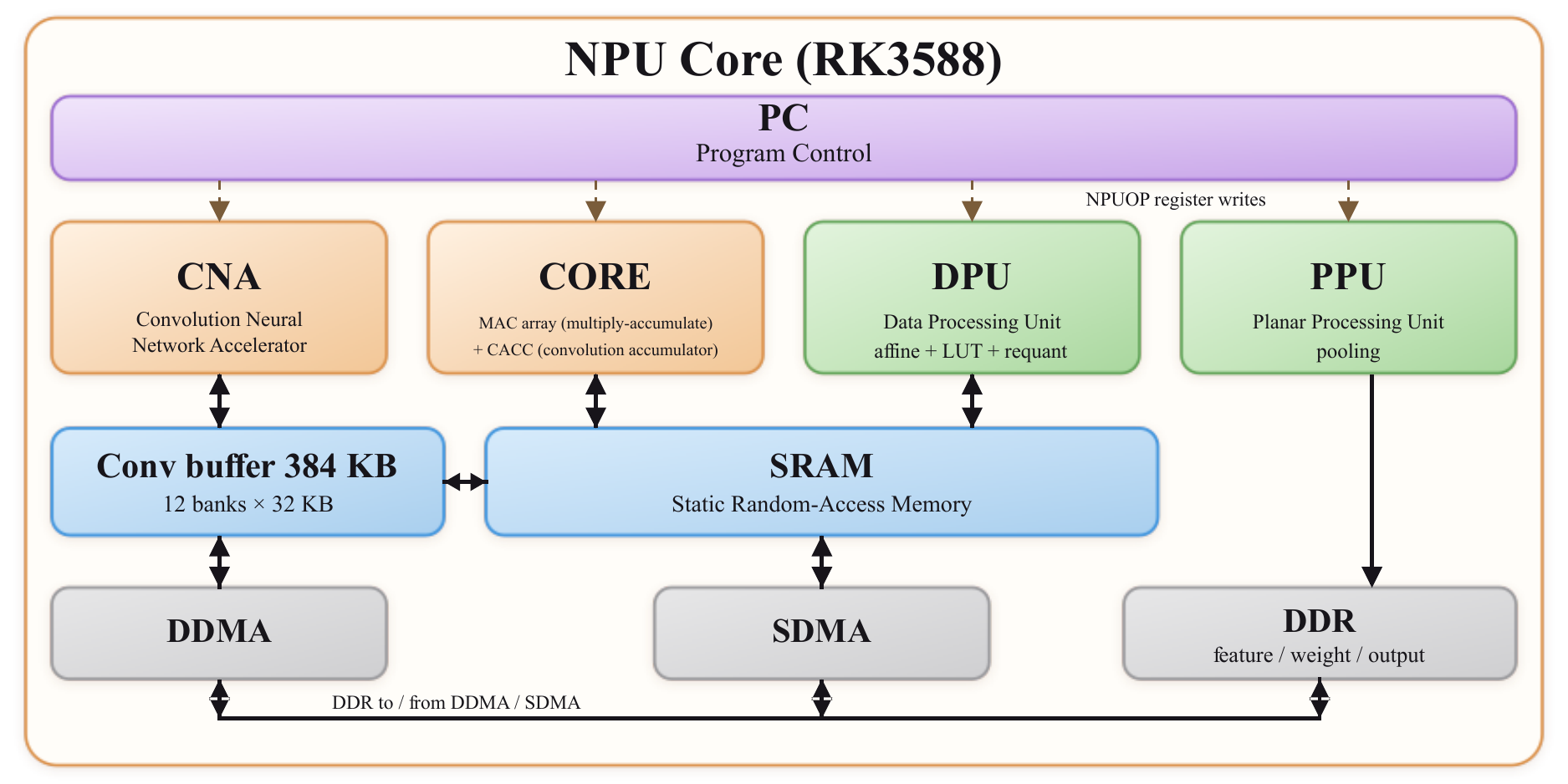}
    \caption{Rockchip RKNPU (RK3588).}
    \label{fig:backends-rk}
  \end{subfigure}
  \caption{On-device inference backends: (a) the Hexagon v79 HTP and (b) the RK3588 NPU core.}
  \label{fig:backends}
\end{figure}

Edge NPUs accelerate neural-network inference through specialized compute units for matrix computation and non-matrix processing~\cite{qualcomm,rk3588_trm,jouppi2020tpu,neuroncore,medina2020gaudi,liao2019davinci}. These units can execute asynchronously, overlapping computation and data movement when hardware capabilities and data dependencies permit. Figure~\ref{fig:backends} illustrates two representative designs: Qualcomm Hexagon~\cite{qualcomm} and Rockchip RKNPU~\cite{rk3588_trm}.

The Hexagon and RKNPU designs shown in Figures~\ref{fig:backends-hx} and~\ref{fig:backends-rk} include hardware units that serve three main roles. {Scheduling} is handled by Hexagon’s scalar Q6 core and RKNPU’s PC task controller, which coordinate their respective execution pipelines~\cite{qualcomm,rk3588_trm}. {Matrix computation} is accelerated by Hexagon’s HMX and RKNPU’s CNA pipeline~\cite{hexagonhmx,rk3588_trm}. {Non-matrix processing} is handled by Hexagon’s programmable HVX vector unit, which supports softmax, normalization, and elementwise operations, and RKNPU’s specialized DPU, which supports elementwise operations and lookup-table-based activations~\cite{qualcomm,rk3588_trm}.


These units can overlap matrix computation, non-matrix processing, and memory transfers when hardware capabilities and data dependencies permit. For example, a matrix engine may process one tile while a non-matrix unit processes results from an earlier tile. On Hexagon, DMA transfers data between main memory and vector tightly coupled memory (VTCM), enabling data movement to overlap with computation~\cite{qualcomm}. Scheduling and synchronization coordinate these operations and ensure that inputs are ready before use. Thus, inference latency depends not only on peak compute throughput, but also on workload distribution across units, memory bandwidth, and the overlap achieved between computation and data movement.


\subsubsection{Attention Execution on NPUs.}
\label{sec:attn-runtime}
\label{sec:async-pipeline}

{Scheduling units dispatch attention tasks; matrix and non-matrix units compute and exchange intermediate results. Below, we outline SDPA, FlashAttention, and GDN execution on RKNPU and Hexagon.}

\textbf{Full attention.} On RKNPU, CNA executes attention matrix products, while DPU supports elementwise processing~\cite{rknnllm,rk3588_trm}. On Hexagon, FlashAttention uses HMX for matrix products and HVX for online softmax and output updates, with computation and data transfers overlapped across tiles when dependencies permit~\cite{geniex,hexagonhmx}.

\textbf{Linear attention.} On Hexagon, the evaluated token-by-token GDN implementation updates recurrent states on HVX, whereas the chunk-based kernel combines HMX GEMMs with HVX processing~\cite{geniex,zhang2026inverse,hexagonhmx}. Recurrent state dependencies constrain overlap between successive updates.

\subsection{On-Device Inference Runtime for VLMs}

{The inference runtime maps model operations to hardware-specific kernels and coordinates execution. Image preprocessing adjusts resolution, strongly shaping the VLM's visual workload.}

\subsubsection{Visual Preprocessing.}
\label{sec:visual-preproc}
During inference, VLMs typically retain the visual preprocessing pipeline used during pretraining to avoid accuracy degradation. Three common strategies \ie fixed-resolution, dynamic-resolution, and tile-based preprocessing, shape the visual workload and inference latency, as detailed below.


{With fixed-resolution preprocessing, images are resized to a preset size, keeping visual-patch and visual-token counts constant across input resolutions.} With {dynamic resolution}, inputs are resized to approximately preserve their original aspect ratios within a predefined pixel or token budget. Larger images yield more patches and visual tokens until the budget is reached~\cite{qwen3.5,diao2026pixels}. With {tile-based preprocessing}, input images are resized to a grid whose tile count is chosen according to their resolution and aspect ratio, then cropped into fixed-size tiles using a multi-crop procedure. A downscaled global thumbnail may also be included. The visual sequence length remains constant for a fixed tile count and increases as additional tiles are introduced~\cite{internvl35,11093135}.

\subsubsection{Edge Inference Frameworks.} Edge inference frameworks enable the full VLM pipeline, including visual embedding generation, multimodal prefill, and autoregressive decoding, on specific hardware platforms. State-of-the-art frameworks for edge devices include llama.cpp~\cite{llamacpp}, MNN~\cite{alibaba2020mnn}, GenieX~\cite{geniex}, and RKLLM~\cite{rknnllm}.




{llama.cpp} is a general-purpose inference framework that supports CPU, GPU, and Qualcomm NPU backends through ggml~\cite{llamacpp}. {MNN} provides highly optimized inference implementations for Arm CPUs and mobile GPUs~\cite{alibaba2020mnn}. {GenieX} is a Qualcomm runtime optimized for the Hexagon NPU, mapping model graphs to hardware-specific implementations~\cite{geniex}. {RKLLM} is an inference stack optimized for the RKNPU~\cite{rknnllm}.

In the following sections, we evaluate inference performance using MNN for its state-of-the-art on-device CPU/GPU execution~\cite{alibaba2020mnn}, and GenieX and RKLLM for their hardware-specific NPU optimizations targeting Qualcomm Hexagon and Rockchip RKNPU, respectively~\cite{geniex,rknnllm}.


\section{Methodology}
\label{sec:methodology}

To answer RQ1 and RQ2, we evaluate diverse model designs and compare NPU execution with CPU and GPU baselines. To answer RQ3, we combine resolution-dependent task accuracy with latency measurements. Accordingly, \S~\ref{sec:models} defines the model and task coverage, \S~\ref{sec:runtime-hardware} specifies the deployment platforms and inference configurations, and \S~\ref{sec:pipeline} establishes shared protocols for accuracy and performance measurements.

\subsection{Models and Tasks}
\label{sec:models}

This subsection defines the six VLMs under test, summarized in Table~\ref{tab:models}, and the ten evaluation benchmarks with their scoring metrics.

\subsubsection{Models}
\label{sec:model-list}

\begin{table}[t]
  \centering
  \caption{The six VLMs under test. Full Attention denotes softmax-based attention; counts following $\times$ give backbone layers. Preprocessing categories follow the default \texttt{huggingface/transformers} configuration for square inputs~\cite{wolf-etal-2020-transformers}.}
  \label{tab:models}
  \scriptsize
  \setlength{\tabcolsep}{3pt}
  \begin{tabularx}{\linewidth}{@{}llrrl>{\raggedright\arraybackslash}X@{}}
    \toprule
    Model & Architecture & Encoder Size (M) & Backbone Size (B) & Attention & Preprocessing \\
    \midrule
    Qwen3.5-2B       & Encoder-based & 331.417 & 1.943 & Full Attention $\times6$ + Linear Attention $\times18$ & Dynamic \\
    Gemma4-E2B       & Encoder-based & 168.545 & 4.647 & Full Attention $\times35$ & Fixed \\
    InternVL3.5-1B   & Encoder-based & 309.265 & 0.752 & Full Attention $\times28$ & Tile-based \\
    SmolVLM2-500M    & Encoder-based & 98.230  & 0.409 & Full Attention $\times32$ & Fixed \\
    NEO1.5-2B        & Encoder-free  & 9.178   & 2.887 & Full Attention $\times40$ & Dynamic \\
    Mono-InternVL-2B & Encoder-free  & 14.249  & 3.097 & Full Attention $\times24$ & Tile-based \\
    \bottomrule
  \end{tabularx}
\end{table}


To study widely used VLMs suitable for edge deployment, we select six representative models with nominal sizes of roughly 500M--3B~\cite{qwen3.5,gemma4,internvl35,smolvlm,diao2026pixels,11093135}, spanning different visual-preprocessing strategies, model architectures, and attention architectures. Table~\ref{tab:models} summarizes the selected models. We deploy each model using the inference framework and quantization profile supported by each device. For encoder-free models, the encoder parameter count includes visual embeddings and input-side projections.

\subsubsection{Tasks and Metrics}
\label{sec:tasks}
We evaluate both model inference latency and accuracy across a broad range of tasks using 10 widely used multimodal benchmarks from \texttt{lmms-eval}~\cite{zhang-etal-2025-lmms}, listed below.

{\textbf{Tasks.} The 47{,}374 labeled questions span multimodal reasoning (MMMU~\cite{10656299}), perception (MMStar~\cite{NEURIPS2024_2f8ee6a3}), science QA (SQA-Img, ScienceQA's image-only subset~\cite{NEURIPS2022_11332b6b}), real-world QA (RealWorldQA~\cite{xai2024realworldqa}), text recognition (OCRBench~\cite{liu2024ocrbench}), mathematical reasoning (MathKangaroo~\cite{10.1007/978-3-032-39150-6_4}), multilingual text-centric QA (MTVQA~\cite{tang-etal-2025-mtvqa}), web-page understanding (VisualWebBench~\cite{visualwebbench}), spatial understanding (ViewSpatial~\cite{10.1007/978-3-032-37098-3_6}), and scene understanding (MME-RealWorld~\cite{ICLR2025_df29d63a}).}

\textbf{Metrics.} We follow the standard evaluation protocol for each benchmark. OCRBench uses its official 0--100 aggregate score. VisualWebBench takes the unweighted mean of seven subset scores, using accuracy, ROUGE-L, or F1 as appropriate. MME-RealWorld reports sample-weighted accuracy (Avg) and the unweighted mean accuracy across subtasks (Avg-C).

\subsection{Hardware and Inference Frameworks}
\label{sec:runtime-hardware}

Next, we introduce the devices, inference frameworks, and quantization settings used in our measurement study.

\subsubsection{Devices}
\label{sec:devices}
\label{sec:hardware}

\begin{table}[t]
  \centering
  \caption{Hardware used in the measurement. VTCM denotes vector tightly coupled memory. Hexagon HTP v79 unit counts reflect our measurements on the OnePlus 13T.}
  \label{tab:devices}
  \footnotesize
  \setlength{\tabcolsep}{4pt}
  \newcolumntype{L}{>{\raggedright\arraybackslash}X}
  \begin{tabularx}{\linewidth}{@{}p{0.07\linewidth}LLL@{}}
    \toprule
                   & OnePlus 13T        & Rhino Pi-X1         & Orange Pi 5 Plus \\
    \midrule
    SoC            & Qualcomm SM8750   & Qualcomm QCS8550    & Rockchip RK3588 \\
    CPU            & 2$\times$Oryon Prime + 6$\times$Oryon Perf. & 1$\times$GoldPlus + 2$\times$Gold + 3$\times$Silver & 4$\times$A76 + 4$\times$A55 \\
    GPU            & Adreno 830         & Adreno 740          & Mali-G610 MP4 \\
    NPU            & Hexagon HTP v79\newline (Q6 scalar + 6 HVX + 1 HMX)\newline VTCM 8\,MiB & Hexagon HTP v73\newline (Q6 scalar + 4 HVX + 1 HMX)\newline VTCM 8\,MiB & RKNPU\newline (CNA + DPU + PPU)$\times$3 \\
    RAM            & 16\,GB             & 16\,GB              & 16\,GB \\
    \bottomrule
  \end{tabularx}
\end{table}

To study how NPU architecture and deployment platform affect
on-device VLM inference, we select three platforms spanning two
generations of Qualcomm Hexagon NPUs and an alternative NPU
architecture, as summarized in Table~\ref{tab:devices}.
Qualcomm's Snapdragon 8 series is a representative platform for
on-device AI, a market in which
Qualcomm accounted for 58\% of global SoC shipments in
2025~\cite{counterpoint2026premiumsoc}. Within this family,
we select Hexagon v73, used in Snapdragon 8 Gen 2 and the embedded
QCS8550 platform, and Hexagon v79, used in Snapdragon 8
Elite~\cite{qualcomm_qairt,qcs8550_brief}.
The Hexagon v73 configuration combines Q6 scalar control, four programmable
HVX vector units, and one HMX matrix accelerator, with 8\,MiB of
VTCM supporting on-chip data reuse.
The Hexagon v79 configuration increases the HVX count
from four to six, while retaining one HMX unit and 8\,MiB of VTCM,
as summarized in Table~\ref{tab:devices}.
Selecting both therefore captures an established Hexagon configuration
and its subsequent expansion of vector resources across smartphone
and embedded deployments.
To complement this generational comparison, we include the
RK3588-based Orange Pi 5 Plus, whose RKNPU uses specialized
CNA, DPU, and PPU units~\cite{rk3588_trm}.

\subsubsection{Inference Frameworks and Quantization}
\label{sec:runtime}

To evaluate on-device inference on CPUs, GPUs, and NPUs, we select the best-performing inference framework for each backend in 2026.

\textbf{Inference frameworks.} We use MNN 3.6.1~\cite{alibaba2020mnn} for CPU and OpenCL GPU baselines, GenieX 0.7.0~\cite{geniex} with ggml-hexagon for Qualcomm Hexagon, and RKLLM 1.3.1~\cite{rknnllm} for Rockchip RKNPU. We ensure that all operators in the evaluated models run on the target hardware, \ie the NPU or GPU, without CPU fallback.


\textbf{Quantization.} For MNN, we use INT8 for the vision module and 4-bit HQQ for the LLM module~\cite{badri2023hqq,alibaba2020mnn}. For GenieX, we use the GGUF weight format~\cite{llamacpp}, with Q4\_0 for the LLM module and Q8\_0 for the vision module, except for NEO1.5's convolutional vision module, which remains in F16. For RKLLM, we use INT8 for both weights and activations~\cite{rknnllm}.




\subsection{Measurement Pipeline}
\label{sec:pipeline}

\begin{wrapfigure}[13]{R}{0.5\linewidth}
  \centering
  \vspace{-10pt}
  \includegraphics[width=\linewidth]{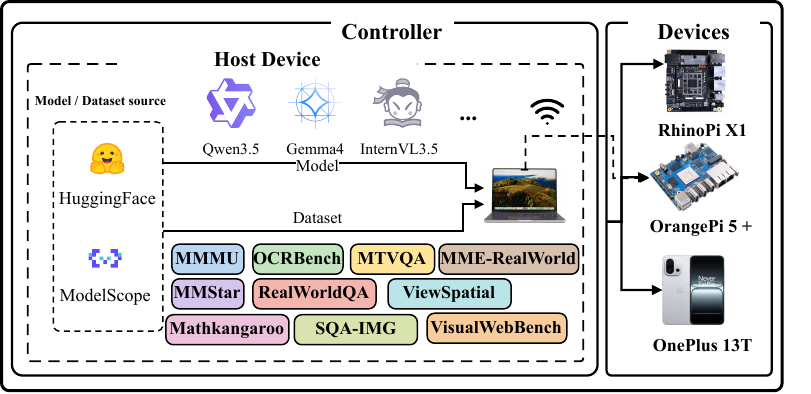}
  \caption{Measurement pipeline: the host fetches models and datasets, deploys them via SSH/ADB, and collects inference outputs and runtime measurements.}
  \label{fig:pipeline}
\end{wrapfigure}
We measure latency, memory usage, and thermal behaviour to assess on-device VLM efficiency, and evaluate accuracy on public benchmarks using the same deployment configurations. All evaluations use the setup in Figure~\ref{fig:pipeline}, comprising a controller host and the three devices in Table~\ref{tab:devices}.

For latency measurements, we configure each backend to run at its highest target frequency: CPUs at 4.32/3.53\,GHz on OnePlus 13T, 3.2/2.8/2.0\,GHz on Rhino Pi-X1, and 2.304/1.8\,GHz on Orange Pi 5 Plus; the Rhino GPU at 680\,MHz; and the Orange Pi GPU/NPU at 1.0\,GHz. We use synchronous single-request execution, with 2\,s between runs. Each process loads one model, and the measurement excludes process startup and model loading.


We measure vision-encoding throughput (megapixels/s, MP/s), prefill throughput (input tokens/s), and decode throughput (output tokens/s) in separate tests, varying image resolution, input sequence length, and initial context length, respectively. Vision-encoding throughput is the input image's megapixel count divided by its encoding latency. Sequence and context lengths are based on actual token counts. We monitor memory usage during prefill by sampling process and system counters through Linux procfs, and thermal behaviour during inference by sampling temperatures through Linux thermal sysfs~\cite{linux_kernel_docs}, both at a target interval of 100~ms. On the OnePlus 13T, \emph{NPU-zone} temperature is the maximum reading across the HVX/HMX thermal sensors at each sampling point, while \emph{skin-zone} temperature represents the device's exterior surface temperature~\cite{androidthermalmitigation} and is reported as the maximum reading across the selected skin thermal sensors. For each configuration, we perform three warm-up runs followed by ten measured runs and report the mean of the measured results.

\section{VLM Inference Performance on Edge NPUs at a Glance}
\label{sec:overall-performance}

\begin{figure*}[!htbp]
  \centering
  \includegraphics[width=\linewidth]{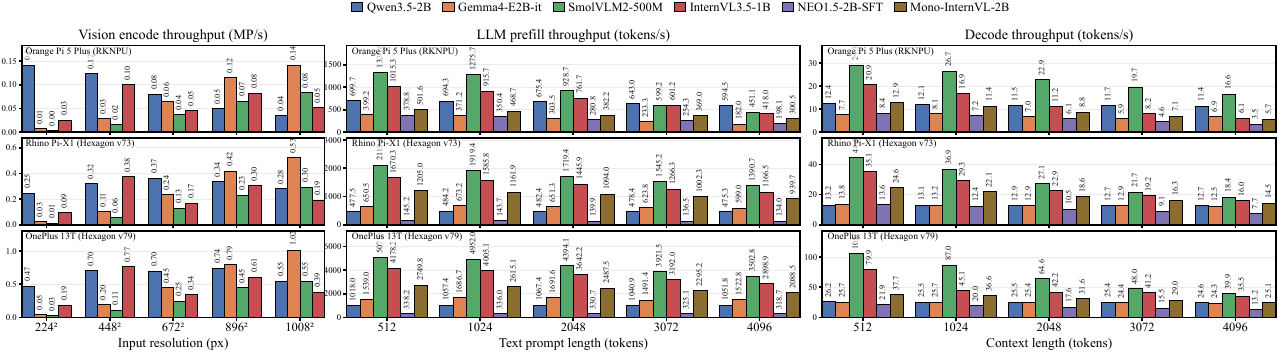}
  \caption{Vision-encoding, prefill, and decode throughput of six VLMs across three NPU stacks. Vision inputs: 224--1008 pixels per side; prefill inputs and initial decode contexts: 512--4096 tokens. Colours identify models; RKLLM and GenieX use RKNPU and Hexagon, respectively.}
  \label{fig:stage-performance-throughput}
\end{figure*}

\begin{figure*}[!htbp]
  \centering
  \includegraphics[width=\linewidth]{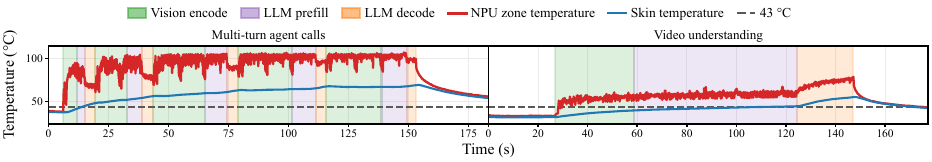}
  \caption{NPU-zone and skin-zone temperatures of Qwen3.5-2B on the OnePlus 13T with GenieX during five inference turns in the MAI-UI Wi-Fi navigation task with 1080$\times$2400 screenshots~\cite{maiuinavigationdemo} and video understanding on VideoXum with 512 frames at 224$\times$224 and 512 output tokens~\cite{lin2024videoxumcrossmodalvisualtextural}. The dashed line marks a 43$^\circ$C surface-temperature reference~\cite{chen2026thermalexperience}.}
  \label{fig:stage-performance-thermal}
\end{figure*}

\begin{wrapfigure}[15]{R}{0.32\textwidth}
  \centering
  \includegraphics[width=\linewidth]{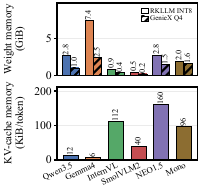}
  \vspace{-20pt}
  \caption{LLM weight sizes (top) and FP16 KV-cache memory per token (bottom).}
  \label{fig:stage-performance-memory}
\end{wrapfigure}

Before presenting the detailed kernel-level performance analysis,
we first examine on-device VLM inference in two real-world workloads:
GUI interaction and video understanding. For both workloads, we measure
NPU inference speed and memory footprint across six VLMs
(Qwen3.5-2B, Gemma4-E2B, InternVL3.5-1B, SmolVLM2-500M,
NEO1.5-2B-SFT, and Mono-InternVL-2B) and characterize thermal
behaviour using Qwen3.5-2B as a representative case.
For GUI interaction, we use the MAI-UI Wi-Fi navigation
example~\cite{maiuinavigationdemo}. This workload comprises five
inference turns, each taking one to three screenshots and a textual
action history as input to generate the next action or a completion
response. For video understanding, we use a video-summarization
workload following the VideoXum input
configuration~\cite{lin2024videoxumcrossmodalvisualtextural},
processing 512 frames at 224$\times$224 resolution sampled at
1~FPS before generating a summary.
Figure~\ref{fig:stage-performance-throughput} reports vision encoding,
prefill, and decode throughput across the six models.
Figure~\ref{fig:stage-performance-thermal} shows temperature traces
during Qwen3.5-2B inference on the OnePlus 13T using GenieX for the
two workloads. Figure~\ref{fig:stage-performance-memory} reports
the deployed LLM weight sizes and KV-cache budgets.

\textbf{Memory footprint.} Current edge devices provide sufficient
unified memory for the tested VLMs, with 12~GB or more of DRAM
becoming common in high-end smartphones~\cite{trendforce2026mobiledram}.
LLM weights and the KV cache are major contributors to the inference
memory footprint.

Figure~\ref{fig:stage-performance-memory} shows the size of each
model's deployed LLM weight file, which depends strongly on
quantization settings. With GenieX Q4\_0, sizes range from 0.2~GiB
for SmolVLM2-500M to 2.5~GiB for Gemma4-E2B. For the same models,
RKLLM INT8 exports are larger, ranging from 0.5~GiB to 7.4~GiB,
respectively.

The KV-cache size depends on context length and per-token memory
usage. Excluding fixed states, FP16 KV-cache usage per token varies
by more than an order of magnitude across models: 6~KiB/token for
Gemma4-E2B, 12~KiB/token for Qwen3.5-2B, and 160~KiB/token for
NEO1.5-2B-SFT, whose 40 full-attention layers each maintain a
4~KiB/token FP16 cache. At a 16K-token context, the growing KV
cache alone occupies approximately 2.5~GiB for NEO1.5-2B-SFT and
1.8~GiB for InternVL3.5-1B, but only 0.1~GiB for Gemma4-E2B.

\textbf{Inference speed.} Figure~\ref{fig:stage-performance-throughput}
reports NPU throughput. {Across three NPUs, Qwen3.5-2B and NEO1.5-2B-SFT average 729 and 254~input tokens/s for prefill, and 16.7 and 11.4~output tokens/s for decoding, respectively.} {The total latency of the workload also depends on visual and textual input length, output length, and the inference turns.} For the Wi-Fi navigation workload, Qwen3.5-2B on the OnePlus 13T
using GenieX takes up to 37.4~s per turn and 146.4~s across the
five inference turns, including vision encoding, prefill, and decoding,
as shown by the stage intervals in Figure~\ref{fig:stage-performance-thermal}.

For the VideoXum workload with 512~input frames and a 512-token
summary, Figure~\ref{fig:stage-performance-thermal} shows that
Qwen3.5-2B on OnePlus 13T using GenieX spends 97.6~s on
vision encoding and prefill before decoding begins and 120.3~s
in total to generate the summary. {For the same video workload}, NEO1.5-2B-SFT takes 153.3~s of model inference on OnePlus 13T.
These results indicate that current NPU throughput remains insufficient for fast responses in these workloads.



\begin{wrapfigure}[13]{R}{0.32\textwidth}
  \centering
  \vspace{-10pt}
  \includegraphics[width=\linewidth]{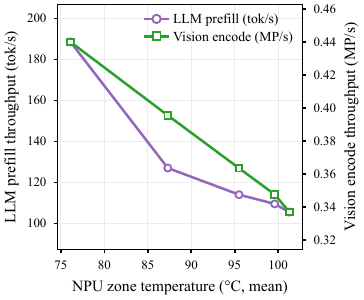}
  \vspace{-20pt}
  \caption{Stage throughput versus mean NPU-zone temperature during MAI-UI.}
  \label{fig:thermal-throughput}
\end{wrapfigure}

\textbf{Thermal behaviour.} We also measure temperature changes
during inference. As shown in Figure~\ref{fig:stage-performance-thermal},
during Qwen3.5-2B inference on the OnePlus 13T's Hexagon v79 NPU
using GenieX for GUI interaction, NPU-zone temperatures rapidly
reach approximately 95$^\circ$C during the first vision-encoding
stage and peak at 107$^\circ$C in subsequent turns, as the
high-resolution GUI screenshots impose substantial vision-encoding
and prefill workloads. In contrast, NPU-zone temperatures remain
mostly around 50--65$^\circ$C while processing the low-resolution
frames for the video understanding task, before rising to
approximately 78$^\circ$C during decoding. Heat generated by
sustained computation spreads to the device body, raising skin-zone
temperatures. Skin-zone temperatures continue to rise across
successive execution phases, peaking at approximately 69$^\circ$C
for the GUI task and 55$^\circ$C for the video understanding task.

This heat accumulation also degrades sustained inference performance. Figure~\ref{fig:thermal-throughput} shows that throughput during vision encoding and prefill decreases as the NPU heats up during the MAI-UI workload. At high temperatures, thermal throttling lowers the NPU's operating frequency, reducing throughput in both stages.


{Our measurements of two on-device VLM applications show that the tested devices have sufficient memory for inference, but latency and thermal management limit deployment. We therefore examine how model architecture, framework runtime, and NPU execution characteristics affect inference efficiency and discuss deployment implications.}

{In \S~\ref{sec:kernel-bottleneck}, we analyze kernel bottlenecks in vision encoding and prefill; in \S~\ref{sec:decode-performance}, we examine how attention affects CPU/NPU decoding efficiency; and in \S~\ref{sec:resolution}, we study model and input resolution selection based on accuracy and latency.}



\section{Vision Encoding and Prefill Efficiency on Edge NPUs}

\label{sec:kernel-bottleneck}
\label{sec:stage-performance}
Prior work identifies the NPU as an efficient accelerator for compute-intensive workloads in on-device inference, including vision encoding and LLM prefill~\cite{cai2026npureadyllmsdissecting,xu2024fastondevicellminference,murthy2026phasematterscharacterizingheterogeneous}. The high latency of vision encoding and prefill measured in \S~\ref{sec:overall-performance} motivates our RQ1: \emph{\rqone}

\begin{figure}[t]
  \centering
  \includegraphics[width=\linewidth]{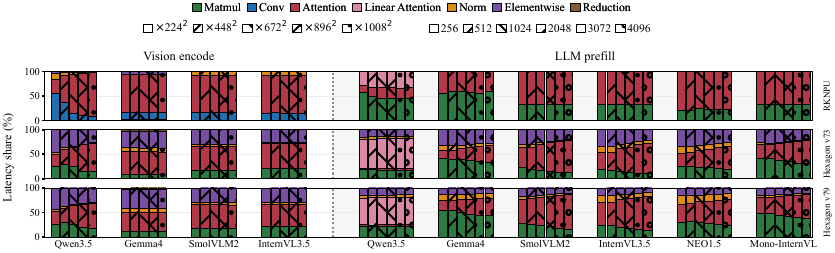}
  \caption{We profile compute-intensive kernel latency shares in vision encoding and LLM prefill under a single request on three NPUs: OnePlus 13T Hexagon v79, Rhino Pi-X1 Hexagon v73, and Orange Pi 5 Plus RKNPU. Each row corresponds to one NPU. The left half reports vision encoding for the four encoder-based models. The right half reports LLM prefill for all six models. We test image resolutions from 224$\times$224 to 1008$\times$1008 and prefill input lengths from 256 to 4096 tokens to cover different workload intensities.}
  \label{fig:kernel-share}
\end{figure}

{Figure~\ref{fig:kernel-share} shows profiled kernel latency by image resolution and prefill length. Full attention contributes up to 89.4\% for Qwen3.5-2B vision encoding on RKNPU and 66.5--70.3\% for SmolVLM2-500M prefill on Hexagon. Linear attention (GDN) contributes 54.5--61.7\% for Qwen3.5-2B prefill on Hexagon v73/v79 over 256--4096 input tokens. Full-attention latency rises with resolution and prefill length; GDN's latency share remains nearly constant over the tested lengths.}

These observations motivate a detailed analysis of the execution of full attention (\S~\ref{sec:full-attention}) and linear attention (\S~\ref{sec:gdn-inefficient}) on edge NPUs.

\subsection{Softmax Limits Full-Attention Efficiency on NPUs}
\label{sec:full-attention}



To identify the factors limiting full-attention efficiency on edge NPUs, we analyze two common implementations: the FlashAttention kernel in GenieX's Hexagon backend~\cite{dao2022flashattentionfastmemoryefficientexact} and the scaled dot-product attention (SDPA) implementation in RKLLM for RKNPU~\cite{vaswani2017attentionisallyouneed}. We find that softmax limits full-attention performance in both implementations. Specifically, it saturates the compute capacity of Hexagon's HVX unit (\S~\ref{sec:fa-softmax-bound}) and forms a bottleneck on RKNPU (\S~\ref{sec:rknpu-softmax-bound}). In \S~\ref{sec:hmx-bubble}, we further examine how the imbalance between softmax and matrix computation leaves NPU matrix engines idle. We also discuss why this underutilization persists during inference with a batch size of one, which is typical on edge devices, and how it can be potentially avoided.


\subsubsection{HVX Limits FlashAttention Performance}
\label{sec:fa-softmax-bound}

On Hexagon NPUs, the Hexagon Matrix eXtensions (HMX) compute multiple dot products in parallel and retain partial sums in dedicated accumulators~\cite{hexagonhmx}. The Hexagon Vector eXtensions (HVX) accelerate elementwise operations, reductions, and data rearrangement using wide single-instruction, multiple-data (SIMD) instructions that process multiple elements in parallel and rearrange them within and across vectors~\cite{hexagonhvx}. In GenieX's FlashAttention implementation, Q6 coordinates data movement and schedules asynchronous execution on HMX and HVX, overlapping their work across tiles. HMX computes attention scores and multiplies attention weights by the values, while HVX computes softmax and writes back completed output tiles in parallel with HMX computation~\cite{geniex}. Data dependencies constrain this overlap: HMX must wait for inputs produced by HVX, and the slower unit limits pipeline throughput. Therefore, we examine which unit limits the performance of the attention kernel.


\begin{figure}[!ht]
  \centering
  \begin{subfigure}{0.344\linewidth}
    \centering
    \includegraphics[width=\linewidth]{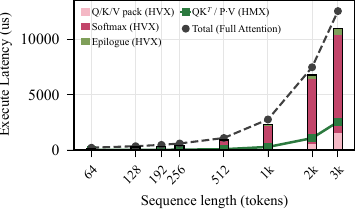}
    \caption{Softmax dominates FA latency}
    \label{fig:fa-bound-a}
  \end{subfigure}\hfill
  \begin{subfigure}{0.266\linewidth}
    \centering
    \includegraphics[width=\linewidth]{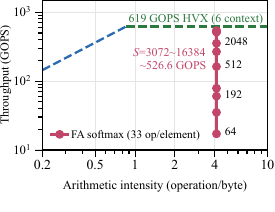}
    \caption{Softmax GOPS roofline}
    \label{fig:fa-bound-b}
  \end{subfigure}\hfill
  \begin{subfigure}{0.305\linewidth}
    \centering
    \includegraphics[width=\linewidth]{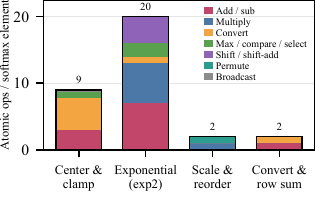}
    \caption{Atomic ops from disassembly}
    \label{fig:fa-bound-c}
  \end{subfigure}
  \caption{FlashAttention on the Hexagon HTP v79. The full kernel is profiled during text prefill as the input sequence length grows from 64 to 3072 tokens, and the softmax used for the roofline is swept over sequence lengths of 64--16384 tokens, with the latency of each point taken as the median of 60 repeated executions.}
  \label{fig:fa-bound-summary}
\end{figure}

\textbf{Softmax execution on HVX is compute-bound, limiting FlashAttention performance.} Figure~\ref{fig:fa-bound-a} shows the latency breakdown of matrix computation, softmax computation, and output processing within a FlashAttention kernel during prefill on Hexagon v79. As shown, Softmax dominates the total kernel latency, particularly at longer input lengths. For an input of 1024 tokens, softmax on HVX takes 1.87~ms out of the 2.76~ms kernel latency, compared with only 0.28~ms for matrix computation on HMX. The proportion of total kernel latency spent on softmax rises from 67.9\% to 71.0\% as the input length increases to 3072 tokens.

Figure~\ref{fig:fa-bound-b} shows the roofline model for HVX on Hexagon v79 alongside softmax throughput. The results clearly show that softmax execution on HVX is compute-bound. Specifically, softmax throughput increases as the sequence length grows from 64 to 3072 tokens, then plateaus at approximately 526.6~GOPS for sequence lengths up to 16384 tokens, reaching 85\% of the 619~GOPS HVX compute ceiling. Because HMX depends on the softmax output to compute the attention output, compute-bound softmax execution on HVX limits overall FlashAttention performance despite HMX's high compute capacity.




\textbf{The lack of native exponential instructions makes HVX the bottleneck in FlashAttention.} Equation~\ref{eq:fa-softmax} gives the softmax formulation:
\begin{equation}
\mathbf{p} = \frac{\exp(\mathbf{s}-\max(\mathbf{s}))}{\operatorname{sum}\!\left(\exp(\mathbf{s}-\max(\mathbf{s}))\right)},
\label{eq:fa-softmax}
\end{equation}
where $\mathbf{s}$ denotes a row of attention scores and $\mathbf{p}$ denotes the corresponding normalized attention weights. Softmax subtracts the row maximum from each score, computes the exponential elementwise, and divides each result by the sum of the exponentials.

{HVX lacks native exponential instructions, so GenieX implements exp2 using vector instructions, which expresses the input as $x=k+r$, where $k$ is an integer and $r$ is the fractional part, so $2^x=2^k2^r\approx2^kP(r)$. {Starting from $a_6$}, it evaluates $P(r)=a_0+a_1r+\cdots+a_6r^6$ by repeatedly multiplying by $r$ and adding the next coefficient, using six multiplications and six additions. It then adjusts the FP16 exponent bits to scale the result by $2^k$. Figure~\ref{fig:fa-bound-c} shows our disassembly results: polynomial evaluation and exponent adjustment account for 20 of the 33 operations per softmax element. For longer sequences, these repeated operations push HVX toward its compute limit, making softmax the bottleneck in FlashAttention.}


\finding[find:fa-softmax-bound]{{Softmax is the execution bottleneck for FlashAttention on Hexagon v79. It drives softmax to 85\% of the measured HVX compute roof, making HVX compute capacity limit FlashAttention performance. }}

\subsubsection{From Softmax Bottlenecks to NPU Pipeline Bubbles}
\label{sec:rknpu-softmax-bound}
\label{sec:hmx-bubble}
Prior work has explored accelerating softmax on CPUs~\cite{hao2025scalingllmtesttimecompute} and dedicated compute units~\cite{rk3588_trm} by replacing expensive exponential computation with lookup table (LUT) operations~\cite{intattention,exaq,cmsis_nn_softmax}. However, we find that reducing the cost of exponential computation alone does not necessarily eliminate the end-to-end throughput bottleneck in full attention. We use RKNPU as a case study and analyze its attention latency breakdown, focusing on softmax execution using the hardware LUT built into its DPU~\cite{rk3588_trm}.



\begin{wrapfigure}[12]{R}{0.32\textwidth}
  \centering
  \vspace{-10pt}
  \includegraphics[width=\linewidth]{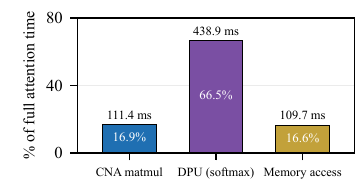}
  \caption{Full-attention step time share on the RK3588 RKNPU~\cite{rk3588_trm} by unit: CNA matmul, DPU (softmax), and memory reads and writes.}
  \label{fig:rknpu-task-granularity}
\end{wrapfigure}

Figure~\ref{fig:rknpu-task-granularity} shows the latency breakdown of full attention on RKNPU. The CNA (Convolution Neural Network Accelerator) performs matrix computation, while the DPU (Data Processing Unit) performs table lookups for softmax computation. Despite using a dedicated compute unit, softmax remains the bottleneck in attention execution. Specifically, softmax computation on the DPU accounts for 66.5\% of the step latency, compared with 16.9\% for matrix multiplications on the CNA.

More importantly, the workload imbalance between the matrix and non-matrix compute units limits end-to-end attention throughput in the asynchronous execution pipeline. Softmax computation can delay the inputs required by the matrix engine. If no independent matrix work is ready, the engine remains idle while waiting for these inputs. We refer to this dependency-induced idle interval as a \emph{pipeline bubble}. Such \emph{bubbles} can also arise on other NPUs, \ie Hexagon with separate matrix and non-matrix execution units~\cite{jouppi2020tpu,neuroncore,medina2020gaudi,liu2016cambricon,liao2019davinci,liao2021ascend,rico2024xdna,masattention}.

\begin{figure}[!htbp]
  \centering
  \includegraphics[width=0.95\linewidth]{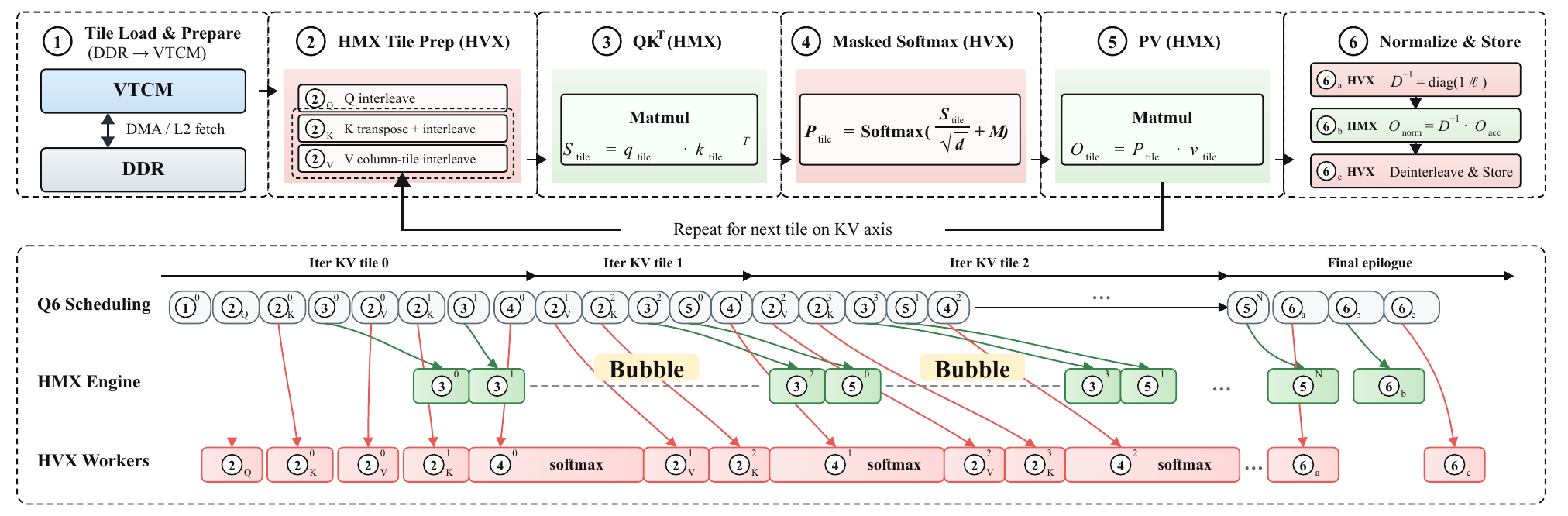}
  \caption{{FlashAttention pipeline on the Hexagon NPU. Top: six execution stages. Bottom: a schematic schedule showing Q6 dispatch and overlapping HMX/HVX execution across successive KV tiles. HMX completes its ready matrix tasks and then waits for HVX softmax outputs, creating pipeline bubbles.}}
  \label{fig:hmx-bubble}
\end{figure}


Figure~\ref{fig:hmx-bubble} illustrates the deployed Hexagon FlashAttention pipeline. Q6 schedules HMX matrix computation and HVX vector computation asynchronously, allowing execution to overlap across tiles. HMX completes its ready matrix tasks before HVX finishes softmax and produces the inputs for the next dependent matrix task. Operand preparation for another tile is also queued behind the same HVX work. HMX therefore remains idle, creating a \emph{pipeline bubble} that recurs across tiles.

Pipeline bubbles pose a unique challenge for edge VLM inference. Cloud serving systems commonly use request- or iteration-level batching to improve hardware utilization~\cite{vllm,orca,sarathiserve}. Multiple concurrent requests provide independent work that can help hide \emph{bubbles} when the runtime permits overlap. Edge inference, however, typically uses a small batch size, often just one request. Optimizations developed for cloud serving may therefore not transfer directly to edge inference.

Our observations suggest that future edge NPU designs could benefit from dedicated softmax acceleration and better workload balance between matrix and non-matrix compute units to improve end-to-end attention performance.



\findingCB[find:rknpu-softmax-bound]{Lookup-table softmax relieves the NPU's attention bottleneck and compute pressure~\cite{hao2025scalingllmtesttimecompute}.}{{Hardware LUT softmax alone cannot eliminate NPU pipeline bubbles. Workload imbalance between matrix and non-matrix units causes idle stalls: on RKNPU DPU LUT-based softmax dominates latency, while on Hexagon HMX waits for HVX softmax results. These bubbles waste matrix-engine resources and persist for single-request edge inference, highlighting the need for balanced throughput between softmax and matrix computation for high attention efficiency.}}

\subsection{Inefficient Linear Attention on Edge NPU}
\label{sec:gdn-inefficient}


{Linear attention~\cite{lineartransformer} represents token history with a fixed-size recurrent state matrix, avoiding a growing KV cache and improving memory and computational scaling. This makes it promising for edge devices with limited memory and compute. However, we find that these algorithmic gains might not always improve NPU performance.}


\begin{figure}[!htbp]
  \centering
  \begin{subfigure}{0.32\linewidth}
    \centering
    \includegraphics[width=\linewidth]{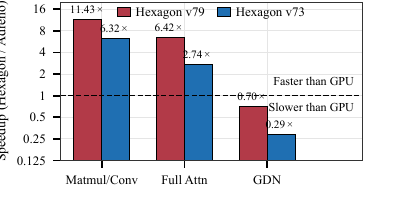}
    \caption{Operator speedup}
    \label{fig:gdn-speedup}
  \end{subfigure}\hfill
  \begin{subfigure}{0.32\linewidth}
    \centering
    \includegraphics[width=\linewidth]{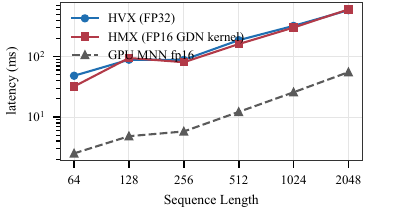}
    \caption{GDN latency}
    \label{fig:gdn-latency}
  \end{subfigure}\hfill
  \begin{subfigure}{0.32\linewidth}
    \centering
    \includegraphics[width=\linewidth]{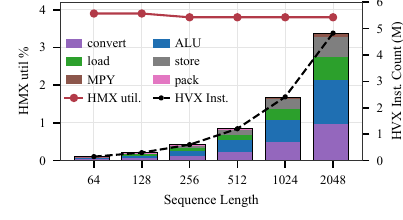}
    \caption{HMX/HVX profiling}
    \label{fig:gdn-hmx-util}
  \end{subfigure}
  \caption{GDN text prefill on Hexagon. Adreno is the GPU baseline; (c) profiles the FP16 HMX kernel.}
  \label{fig:gdn-summary}
\end{figure}

\subsubsection{Linear Attention Underperforms on Hexagon}
\label{sec:gdn-dominates}
We study linear-attention execution on NPUs using GenieX's Gated DeltaNet (GDN) as a representative implementation~\cite{yang2025gdn}. On Hexagon, the compiled GDN kernels execute either token-by-token vector recurrence on HVX or chunked GEMM-based recurrence on HMX. Specifically, the FP32 implementation uses token-by-token vector recurrence, whereas the FP16 implementation uses chunked GEMM-based recurrence~\cite{zhang2026inverse} with a multiplication-only approximation. We evaluate both implementations and use linear-attention performance on Adreno mobile GPUs as a baseline for comparison. Figure~\ref{fig:gdn-summary} illustrates the results.


At a sequence length of 1024~tokens, Figure~\ref{fig:gdn-speedup}
shows that Hexagon NPUs accelerate matrix multiplication and
convolution by $6.3$--$11.4\times$ and full attention by
$2.7$--$6.4\times$ over the Adreno GPU baseline. In contrast,
the HVX GDN kernel delivers approximately 30\% and 71\% lower
throughput on Hexagon v79 and v73, respectively, relative to
MNN OpenCL's chunked GEMM-based GDN implementation on Adreno.

Figure~\ref{fig:gdn-latency} further shows that the FP16 GDN
kernel on HMX still cannot close the performance gap. Across
sequence lengths of 64--2048~tokens, its latency is approximately
990\% to 1867\% higher than that of MNN OpenCL's FP16 GDN
implementation on the Adreno GPU.

\findingCB[find:gdn-prefill]{GDN's fixed-size state should make its attention layers more efficient than full attention in long contexts~\cite{yang2025gdn}.}{On the Hexagon NPUs, GDN runs slower than the Adreno GPU even though full attention is accelerated: at a sequence length of 1024 tokens, GDN achieves only $0.29$--$0.70\times$ speedup relative to the GPU, versus $2.7$--$6.4\times$ for full attention, as Figure~\ref{fig:gdn-speedup} shows. The FP16 HMX kernel stays close to the FP32 HVX recurrence, and both exceed GPU latency across sequence lengths of 64--2048 tokens, as Figure~\ref{fig:gdn-latency} shows.}

\subsubsection{Tensor layout transformations on HVX limit GDN performance on NPUs.}
\label{sec:gdn-misses-hmx}
\begin{figure}[t]
  \centering
  \includegraphics[width=\linewidth]{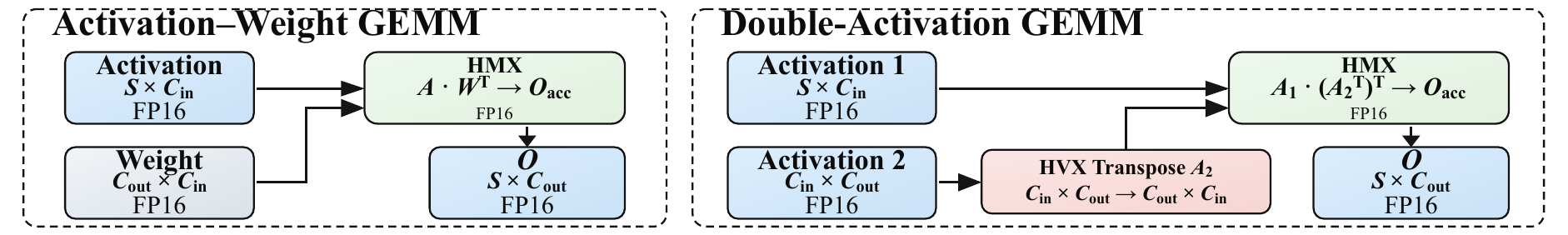}
  \caption{Two HMX GEMM data flows. Left: activation--weight GEMM $A\cdot W^{\top}\to O$ with fixed model weights that can be packed offline and reused. Right: GDN double-activation GEMM $A_1\cdot A_2\to O$, where both operands are activations; because the HMX transposes its second operand automatically, $A_2$ must be pre-transposed by the HVX from $C_{\mathrm{in}}\times C_{\mathrm{out}}$ to $C_{\mathrm{out}}\times C_{\mathrm{in}}$ before every GEMM. Shapes are shown at the paper-level FP16 algorithm view.}
  \label{fig:hmx-dataflow}
\end{figure}

Both the HMX kernel on the NPU and the OpenCL kernel on the GPU use chunked GEMM computations, yet their GDN latencies differ substantially. To explain this gap, Figure~\ref{fig:hmx-dataflow} contrasts a GEMM using model weights that remain fixed during inference with GDN's double-activation GEMMs, both mapped to HMX. The runtime operands of these double-activation GEMMs require additional tensor layout transformations on HVX.


{For a GEMM that maps $M$ input vectors of length $C_{\mathrm{in}}$ to $M$ output vectors of length $C_{\mathrm{out}}$, HMX expects operand shapes of $M\times C_{\mathrm{in}}$ and $C_{\mathrm{out}}\times C_{\mathrm{in}}$~\cite{hexagonhmx}. In a standard linear layer, these input vectors are the token representations processed together. The model weights remain fixed during inference, so they can be converted and packed offline, then reused during inference, as illustrated in Figure~\ref{fig:hmx-dataflow} (left). However, GDN's chunked recurrence introduces many double-activation GEMMs, as shown in Figure~\ref{fig:hmx-dataflow} (right). Activations for later chunks depend on outputs from earlier chunks, so $A_2$ cannot be prepared offline. Instead, HVX must transpose $A_2$ from $C_{\mathrm{in}}\times C_{\mathrm{out}}$ to $C_{\mathrm{out}}\times C_{\mathrm{in}}$ at runtime.}
{As a result, HMX utilization is only 3.8--3.9\%, while HVX instruction counts rise roughly linearly with sequence length and are dominated by conversion, ALU, and load/store operations, as demonstrated in Figure~\ref{fig:gdn-hmx-util}.}

\finding[find:gdn-hmx-incompatible]{{The GDN HMX kernel on Hexagon NPU suffers from HVX-side bottlenecking. Although the chunked matrix representation is offloaded to HMX, HVX is forced to transpose and repack the second operand ahead of each GEMM invocation. Consequently, HMX utilization falls to just 3.8–3.9\%, and HVX overhead dominates kernel execution time (Figure~\ref{fig:gdn-hmx-util}).}\par\smallskip
\textbf{Opportunity:} Where the recurrence's data dependencies {allow}, asynchronous HMX/HVX execution can overlap HVX transposition for one activation operand with HMX GEMM computation on another.}

\section{Decoding Efficiency on Edge NPUs}
\label{sec:decode-performance}




Prefill processes input tokens in parallel using matrix--matrix multiplications (GEMMs), whereas decoding with a batch size of one generates tokens sequentially through repeated matrix--vector multiplications (GEMVs). Each decoding step accesses model weights and attention states, generating substantial memory traffic and making memory bandwidth a key performance constraint on edge devices. Therefore, prior work favors CPUs over GPUs or NPUs for such memory-bound workloads on edge devices~\cite{cai2026npureadyllmsdissecting,xiao2026understandinglargelanguagemodels,zhang2025challenginggpudominancecpus}.

\begin{figure}[!htbp]
  \centering
  \includegraphics[width=\linewidth]{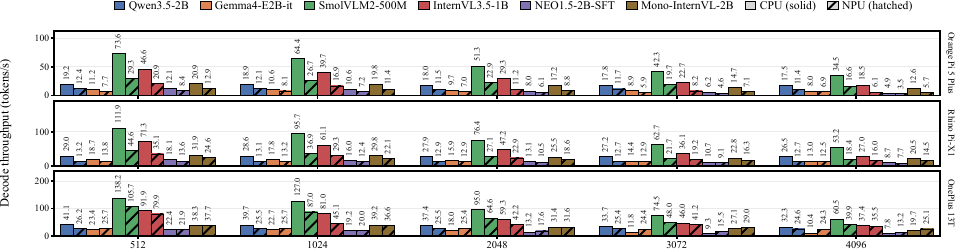}
  \caption{Decode-only throughput over 32 output tokens after text-only prefill with 512--4096-token initial contexts (linear scale).}
  \label{fig:stage-performance-decode}
\end{figure}

However, our results show that neither the CPU nor the NPU consistently provides faster decoding across models, as shown in Figure~\ref{fig:stage-performance-decode}. This motivates RQ2: \emph{\rqtwo} In \S~\ref{sec:decode-attn-latency}, we use module-level CPU/NPU profiles to explain why the faster backend varies across models. In \S~\ref{sec:decode-shape}, we analyze how attention configurations affect execution efficiency on the two backends.



\subsection{CPU Does Not Always Lead in Decoding}
\label{sec:decode-attn-latency}


{Figure~\ref{fig:stage-performance-decode} shows that neither backend consistently decodes faster across the tested deployments. With the context length of 4096, SmolVLM2-500M achieves 60.5~tokens/s on two Oryon Prime CPU cores using MNN while 39.9~tokens/s on the Hexagon NPU using GenieX. For Gemma4-E2B, the NPU is faster: 24.3~tokens/s and 10.4~tokens/s on the CPU.}

\begin{figure}[t]
  \centering
  \includegraphics[width=\linewidth]{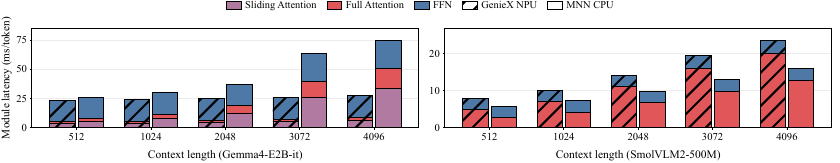}
  \vspace{-20pt}
  \caption{CPU/NPU attention and FFN latency per decode token on OnePlus 13T.}
  \label{fig:decode-backend-a}
   \vspace{-7pt}
\end{figure}

We then profile the attention and FFN components of decoding. Figure~\ref{fig:decode-backend-a} shows the latency of each component summed across all decoder layers per output token. We find that FFN latency varies relatively little with context length and remains comparable between the CPU and NPU, whereas attention latency differs more substantially between the two backends. For example, with a context length of 4096 tokens, SmolVLM2-500M exhibits a CPU/NPU latency gap of 7.4~ms/token for attention but only 0.1~ms/token for FFN. The CPU executes this model's attention faster, whereas the NPU executes Gemma4-E2B's attention faster. These results identify attention execution efficiency as a key factor in decoding performance differences between CPUs and NPUs. {We therefore ask why attention during decoding is slower on NPUs for some models but faster for others. To this end, in \S~\ref{sec:decode-shape}, we examine the execution details of grouped-query attention (GQA) on NPUs, a widely used mechanism that can improve autoregressive decoding efficiency by sharing key and value heads within each group of query heads.}



\subsection{Limited Matrix Parallelism Constrains NPU Attention Efficiency}
\label{sec:decode-shape}



{During autoregressive decoding, the model generates one
output token at each step. Each GQA group contains $G$ query heads
with head dimension $d$ that share a key head and a value head
over a context of $C$ tokens~\cite{ainslie2023gqatraininggeneralizedmultiquery}.}
Full attention combines query--key scoring and value aggregation, with $f$ applying score scaling, masking, and softmax:
\begin{equation}
\mathbf{O}=f(\mathbf{Q}\mathbf{K}^{\top})\,\mathbf{V},
\quad \mathbf{Q}\in\mathbb{R}^{G\times d},
\quad \mathbf{K},\mathbf{V}\in\mathbb{R}^{C\times d},
\label{eq:decode-attention}
\end{equation}
Together, $G$, $d$ and $C$ determine the available matrix parallelism.

\begin{figure}[t]
  \centering
  \includegraphics[width=\linewidth]{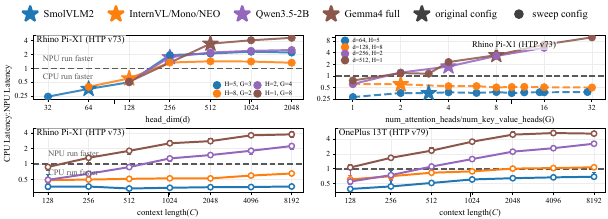}
  \caption{CPU/NPU latency ratio for single-token full attention. Top: $d$ (left) and $G$ (right) sweeps on Rhino Pi-X1 at $C=4096$, holding other head parameters fixed; solid and dashed curves denote HMX and HVX. Bottom: $C$ sweeps on Rhino Pi-X1 (left) and OnePlus 13T (right); InternVL3.5-1B, Mono-InternVL-2B, and NEO1.5-2B share a curve owing to identical attention shapes. Stars and circles mark original shapes and controlled variants. Ratios above one favour the NPU; below one, the CPU.}
  \label{fig:decode-backend-b}
\end{figure}

{Matrix parallelism during decoding comes from both the model's attention architecture and the current context size. The head dimension $d$ and GQA group size $G$ are determined by the model architecture. Figure~\ref{fig:decode-backend-b} shows how these architectural parameters affect HMX-based attention performance relative to the CPU: larger $d$ provides wider matrix operands, while larger $G$ allows more query heads to be processed in parallel using shared K/V data. These properties favour HMX execution in Gemma4-E2B ($d=512$, $G=8$). For a given model, the context size $C$ varies across requests and grows as output tokens are generated; longer contexts provide more K/V tiles to sustain matrix execution. The context sweeps in Figure~\ref{fig:decode-backend-b} show that Qwen3.5-2B's attention is faster on the CPU at short contexts but faster on the NPU at longer contexts, even though its $d$ and $G$ remain unchanged.}

{In contrast, the CPU advantage observed for SmolVLM2, InternVL3.5-1B, Mono-InternVL-2B, and NEO1.5-2B is also tied to GenieX's kernel selection. With $d<192$, the measured GenieX implementation dispatches their decode FlashAttention to HVX~\cite{geniex}. This limits the use of matrix parallelism across $d$, $G$, and $C$, leaving the NPU's HMX matrix acceleration unexploited.}


{These observations suggest an opportunity to dynamically route attention computation during decoding between CPU and NPU backends based on the workload. Workloads with greater matrix parallelism could be routed to the NPU for HMX-accelerated execution, whereas those with limited matrix parallelism or restricted to HVX execution on the NPU could be routed to the CPU. From a kernel-level perspective, selecting the optimal attention backend for autoregressive decoding therefore requires a performance model that quantifies matrix computation efficiency and compares hardware utilization across the CPU and NPU.
}





\findingpairCB[find:decode-attention-shape]
  {The CPU dominates the NPU in decode throughput~\cite{cai2026npureadyllmsdissecting}.}
  {find:decode-attention-cost}{{Neither hardware backend consistently achieves superior decode throughput. Attention efficiency accounts for a large fraction of the CPU–NPU throughput gap, whereas FFN latency remains comparable across the profiled models, as illustrated in Figures~\ref{fig:stage-performance-decode} and~\ref{fig:decode-backend-a}}.}
  {find:decode-gqa-configuration}{{NPU attention efficiency depends on the matrix parallelism  exposed by $d$, $G$, and $C$. Larger head dimensions, groups, and contexts can favour HMX-accelerated NPU execution; small shapes favour the CPU, as shown in Figure~\ref{fig:decode-backend-b}. This performance crossover is contingent on hardware compute capability, memory bandwidth, and runtime‑specific implementation details.} }


\section{Input Resolution as a Lever for Edge VLM Performance}
\label{sec:resolution}

{While \S~\ref{sec:kernel-bottleneck} and \S~\ref{sec:decode-performance} identify kernel-level factors governing inference efficiency on NPUs, this section examines input resolution as an \emph{overlooked lever} that strongly affects end-to-end VLM inference performance on edge devices. In practice, input images are typically processed at the default resolution specified in the model configuration. {However, the accuracy gains from higher resolution depend on the task, while its TTFT cost depends on preprocessing and the inference stack.}}


Motivated by this observation, we pose RQ3: {\emph{\rqthree}} We first examine how resolution affects the visual workload in \S~\ref{sec:resolution-workload}, then define and evaluate task-specific \emph{Sweet Resolutions} in \S~\ref{sec:finding1}.


\subsection{Input Resolution Matters}
\label{sec:resolution-workload}

\begin{wraptable}[15]{R}{0.5\textwidth}
  \centering
  \vspace{-10pt}
  \caption{Visual-token counts for LLM prefill on square inputs (112--1008 pixels); each cell reports the visual-token count. Counts follow each checkpoint's official preprocessing, verified with \texttt{huggingface/transformers}~5.5.4~\cite{wolf-etal-2020-transformers}.}
  \label{tab:visual-tokens}
  \footnotesize
  \setlength{\tabcolsep}{2pt}
  \begin{tabular}{@{}lrrrrrrr@{}}
    \toprule
    & \multicolumn{7}{c}{Visual tokens} \\
    Model & 112 & 224 & 336 & 448 & 672 & 784 & 1008 \\
    \midrule
    Qwen3.5-2B       & 49   & 64   & 100  & 196  & 441  & 576  & 1024 \\
    NEO1.5-2B        & 49   & 64   & 100  & 196  & 441  & 576  & 1024 \\
    Gemma4-E2B       & 256  & 256  & 256  & 256  & 256  & 256  & 256 \\
    SmolVLM2-500M    & 1088 & 1088 & 1088 & 1088 & 1088 & 1088 & 1088 \\
    InternVL3.5-1B   & 256  & 256  & 256  & 256  & 1280 & 1280 & 2560 \\
    Mono-InternVL-2B & 256  & 256  & 256  & 256  & 1280 & 1280 & 2560 \\
    \bottomrule
  \end{tabular}
\end{wraptable}

As described in \S~\ref{sec:visual-preproc}, a VLM's preprocessing typically rescales images to a fixed size, partitions them into tiles, or retains variable dimensions within a pixel budget. These rules determine the images actually processed by the model and map the requested resolution to the visual-patch count for vision encoding and the visual-token count passed to the LLM backbone for prefill. Although the default configuration ensures input compatibility, selecting a resolution for deployment also requires considering the task and target device.

{From the perspective of inference latency, input resolution and the preprocessing strategy largely determine the prefill workload. Table~\ref{tab:visual-tokens} shows how each model's preprocessing strategy maps square single-image inputs to {visual-token counts}. Fixed-size preprocessing, \eg{} {Gemma4-E2B and SmolVLM2-500M}, keeps the visual-token count constant; tile-based preprocessing, \eg{} {InternVL3.5-1B and Mono-InternVL-2B}, increases it at tile boundaries; and dynamic preprocessing, \eg{} {Qwen3.5-2B and NEO1.5-2B-SFT}, {yields visual-token counts that are positively correlated with input resolution}. More specifically, Qwen3.5-2B's visual-token count rises from 64 at {224$\times$224 pixels} to 1024 at {1008$\times$1008 pixels}, a sixteenfold increase. In contrast, InternVL3.5-1B has {1280 visual tokens at both 672$\times$672 and 784$\times$784 pixels}.\par}

{From the perspective of task accuracy, input resolution {can strongly affect} VLM accuracy on specific tasks. For example, high-resolution screenshots for GUI agents and multi-frame video inputs make resolution selection particularly important, as illustrated by the workloads in Figure~\ref{fig:stage-performance-thermal}. Screen understanding requires fine-grained details, such as small text and interface elements, so higher resolution {can be helpful}. In contrast, video understanding involves processing multiple frames and reasoning about causal relationships across them, so increasing {per-frame resolution} may offer less benefit.}

{Input resolution is therefore an important lever for balancing task accuracy and inference latency.}



\subsection{Sweet Resolution}
\label{sec:sweet_resolution}
\label{sec:finding1}


{For a given task and inference stack (model, framework, and hardware), a \emph{Sweet Resolution} refers to a tested input resolution that yields substantial accuracy gains while incurring modest TTFT overhead under a fixed text-prompt length.}

\begin{figure}[!htbp]
  \centering
  \includegraphics[width=\linewidth]{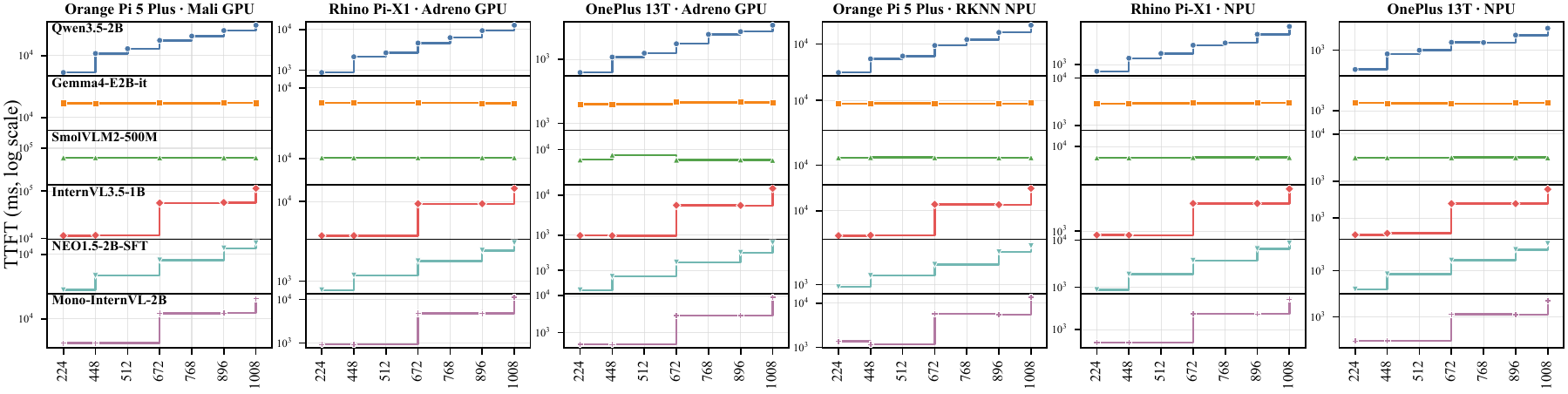}
  \caption{Single-image TTFT versus input resolution for six VLMs across six device$\times$accelerator stacks. The x axis is the resolution in pixels (224--1008), and all measurements use the fixed short prompt ``Describe the image.'' SmolVLM2-500M and Gemma4-E2B stay flat, Qwen3.5-2B and NEO1.5-2B rise continuously, and InternVL3.5-1B and Mono-InternVL-2B exhibit plateaus within each tile-count interval.}
  \label{fig:ttft-staircase}
\end{figure}

\begin{figure}[!htbp]
  \centering
  \includegraphics[width=\linewidth]{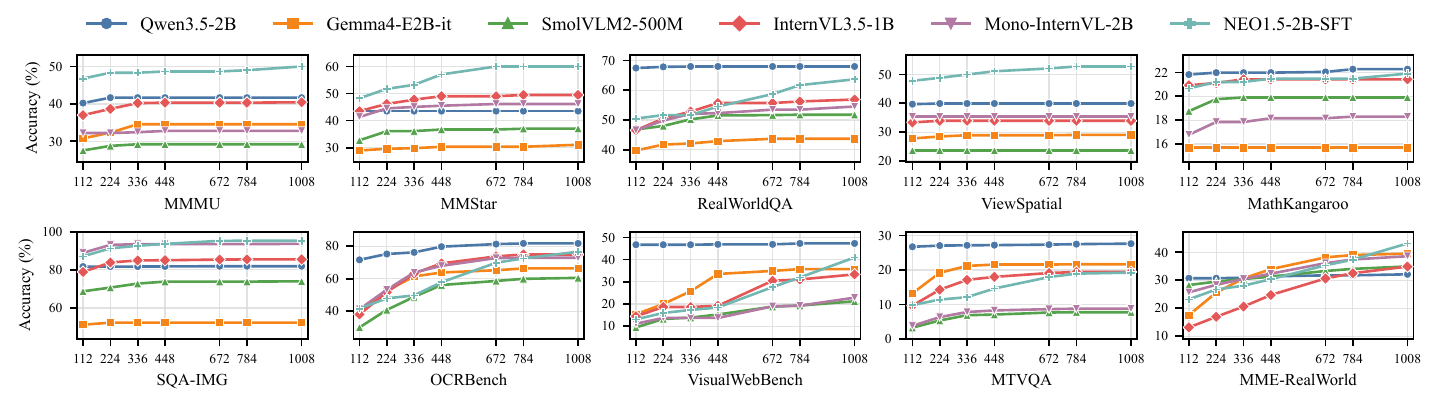}
  \caption{Task accuracy versus input resolution for six VLMs across the ten task categories in \S~\ref{sec:models}, comprising 47{,}374 labeled questions. Each panel corresponds to one task, named on its x axis; the x axis is the input resolution, expressed as the longest edge in pixels (112--1008).}
  \label{fig:resolution-accuracy}
\end{figure}

{To identify sweet resolutions, we evaluate six VLMs with native preprocessing, comparing TTFT across device--backend stacks in Figure~\ref{fig:ttft-staircase} with accuracy on the ten benchmarks of \S~\ref{sec:models} in Figure~\ref{fig:resolution-accuracy}. The visual-token counts in Table~\ref{tab:visual-tokens} help explain the measured TTFT trends. Within a measured TTFT plateau, we retain the tested input with the highest task accuracy; when TTFT rises with resolution, we compare the accuracy gains with the additional TTFT relative to lower-resolution tested inputs on the same inference stack.}

Figure~\ref{fig:ttft-staircase} shows flat, staircase, and continuously rising TTFT curves under fixed, tile-based, and dynamic preprocessing, respectively, consistent with prior work~\cite{Vasu_2025_CVPR,hired}. Figure~\ref{fig:resolution-accuracy} shows that accuracy can improve within a TTFT plateau, particularly on detail-sensitive tasks such as text recognition. Accuracy gains diminish on many tasks, whereas VisualWebBench continues to improve at the largest tested input. 

For example, on MME-RealWorld, InternVL3.5-1B reaches 32.5\% accuracy at 784 pixels, compared to 30.7\% at 672 pixels (Figure~\ref{fig:resolution-accuracy}).
{Table~\ref{tab:visual-tokens} reports identical visual-token counts for 672 and 784 pixels, with a token increase at 1008 pixels. This plateau-and-jump pattern matches the tile-based TTFT staircases shown in Figure~\ref{fig:ttft-staircase}.}
{Treating 784 pixels as the Sweet Resolution, we secure this accuracy gain at only modest TTFT overhead.}
{With fixed preprocessing such as SmolVLM2-500M, the highest-accuracy tested input within the measured TTFT plateau may occur below the maximum resolution. With dynamic preprocessing such as Qwen3.5-2B, TTFT increases with resolution, so candidates are compared by both their accuracy gains and TTFT costs.}

{{These comparisons can guide resolution selection on the target inference stack according to the task's accuracy needs and TTFT budget.}}



\finding[find:sweet-resolution]{
{{A \emph{Sweet Resolution} offers substantial task accuracy gains with modest TTFT overhead. Selection is task-specific and guided by both the accuracy and measured TTFT curves. For fixed- and tile-based preprocessing, the visual-token counts in Table~\ref{tab:visual-tokens} help explain the observed TTFT plateaus; the highest-accuracy tested input within a plateau is a candidate. For dynamic preprocessing, accuracy gains must be considered alongside the increasing TTFT.}}}



\section{Suggestions for Developers}
\label{sec:recommendations}

{Our findings show the mismatches among model architectures, runtime implementations, and NPU capabilities limit on-device VLM inference efficiency. We therefore recommend:}

{\textbf{For hardware:}
\begin{enumerate*}
\item Add dedicated NPU hardware support for softmax exponentiation, balancing throughput with the matrix engine (\S~\ref{sec:full-attention}).
\item Enable flexible HMX operand layouts or hardware-assisted layout conversion to reduce runtime overhead (\S~\ref{sec:gdn-misses-hmx}).
\item {Balance vector and matrix unit throughput on edge NPUs. Mismatches create pipeline bubbles, leaving one unit idle while the other becomes the bottleneck (\S~\ref{sec:hmx-bubble}).}
\end{enumerate*}}

{\textbf{For inference runtime:}
\begin{enumerate*}
\item To mitigate pipeline bubbles, accelerate softmax and overlap independent matrix/vector work where data dependencies permit (\S~\ref{sec:hmx-bubble}).
\item {Select CPU or NPU backends and matrix or vector kernels according to available parallelism to align computation with hardware capabilities and mitigate hardware underutilization.}
\item {{Locate} task-specific Sweet Resolutions, securing meaningful accuracy gains at modest TTFT overhead.}
\end{enumerate*}}

{\textbf{For model architecture:}
\begin{enumerate*}
\item When designing full and linear attention, account for the target edge NPU's operand-layout constraints and available parallelism, rather than relying solely on theoretical algorithmic complexity (\S\S~\ref{sec:gdn-inefficient} and~\ref{sec:full-attention}).
\item Explore attention and normalization designs that replace transcendental operations such as exponentiation with efficient edge-NPU operations. Compare these model-level alternatives with lookup-table implementations for task accuracy and on-device latency (\S~\ref{sec:full-attention}).
\end{enumerate*}}


\section{Related Work}
\label{sec:related-work}

\textbf{On‑device inference benchmarking.}
A rich body of benchmarking work evaluates large‑model inference on heterogeneous mobile hardware. MELT~\cite{laskaridis2024meltingpointmobileevaluation}, Xiao et al.~\cite{xiao2026understandinglargelanguagemodels}, and PowerBench~\cite{cai2026npureadyllmsdissecting} characterize end‑to‑end performance and resource consumption for LLMs across devices and runtimes. Li et al.~\cite{li2026npusfasterstagelevelanalysis}, HeteroInfer~\cite{chen2025characterizingmobilesocaccelerating}, and LLMs-Eval~\cite{zhang2025challenginggpudominancecpus} provide stage‑level NPU analysis and heterogeneous backend‑selection guidance. Phase Matters~\cite{murthy2026phasematterscharacterizingheterogeneous} further profiles VLM phase‑wise behaviour and thermal characteristics on a mobile SoC. Most existing measurements remain at end‑to‑end or stage granularity and lack a kernel‑centric view to expose cross‑layer mismatches between VLM algorithms, runtime kernels, and edge‑NPU hardware. Our work addresses this gap via kernel‑oriented profiling across two distinct NPU architectures (Hexagon and RKNPU), identifying bottlenecks stemming from insufficient cross‑layer coordination.

\textbf{Attention and softmax optimization.}
FlashAttention~\cite{dao2022flashattentionfastmemoryefficientexact}, FlashAttention‑3/4~\cite{fa3,fa4}, and MAS‑Attention~\cite{masattention} optimize attention via tiling, memory reuse, and asynchronous execution, primarily targeting GPUs. For edge‑NPU scenarios, EXAQ~\cite{exaq}, IntAttention~\cite{intattention}, and Hao et al.~\cite{hao2025scalingllmtesttimecompute} propose softmax optimizations including quantization and lookup‑table schemes. HA‑NPU~\cite{hanpu} restructures dataflows to mitigate layout overhead for hybrid‑attention models on NPUs. These works focus on designing new algorithms or custom kernels for performance gains. By contrast, we profile unmodified production inference stacks and uncover real‑world bottlenecks such as softmax‑induced pipeline bubbles and heavy tensor‑layout overhead, which originate from poor cross‑layer coordination between model logic, inference runtime, and NPU hardware units.

\section{Limitations and Future Work}
\label{sec:limitations}

\textbf{Architectural scope and generalizable insights.} This work conducts systematic kernel-centric evaluations on commodity edge NPUs with decoupled matrix and vector execution units, covering representative Hexagon and RKNPU hardware and production-grade inference runtimes. While our key findings regarding compute unit imbalance, pipeline bubbles, and operand layout overheads are grounded in these practical deployment stacks, further validation is required to generalize them to radically different NPU architectures. Within this practical scope, our revealed cross-layer mismatches provide solid architectural insights to guide future hardware and runtime design.

{\textbf{Extending efficiency evaluation for future thermal-aware deployment.} Our evaluation focuses on latency and throughput. The temperature traces in \S~\ref{sec:overall-performance} show heat buildup in selected workloads but do not systematically assess power or thermal efficiency. Future work will measure power draw, energy per inference, and sustained performance under thermal constraints to examine their trade-offs with end-to-end latency.}

\textbf{Foundations for future NPU performance modeling.} Our empirical measurements lay the groundwork for building predictive models that capture how workload shapes, operand layouts, data movement, and cross‑dependencies between matrix and vector computations govern NPU latency and hardware utilization. Future research could develop and validate such performance models across diverse devices and inference runtimes to inform hardware‑architecture design, kernel‑level optimizations, and real‑world deployment choices.


\section{Conclusion}
\label{sec:conclusion}

We empirically characterize on-device VLM inference on commodity edge NPUs and expose cross-layer mismatches across models, runtimes, and hardware. In vision encoding and prefill, softmax dominates execution and triggers matrix-unit pipeline bubbles, which LUT-based softmax cannot fully resolve. Linear attention implementations such as GDN lose their theoretical benefits from costly runtime tensor-layout transformations. For autoregressive decode, CPU and NPU have no universal winner: attention-tensor matrix parallelism decides their relative {throughput. We} discover task-specific Sweet Resolutions that yield substantial task accuracy gains with modest TTFT overhead. We conclude that practical VLM performance depends on model–kernel–hardware mapping, not merely algorithmic complexity, and deliver cross-layer optimization guidance for designers.

\bibliographystyle{ACM-Reference-Format}
\bibliography{references}

\clearpage
\appendix

\section{Appendix}

\subsection{Task Accuracy Under Deployment Quantization}
\label{app:weight-only-quantization}

Following the measurement pipeline in \S~\ref{sec:pipeline}, we evaluate the six models on the ten benchmarks in \S~\ref{sec:tasks} across the target NPUs, using the quantization configurations of their respective inference runtimes in \S~\ref{sec:runtime} and each model's default visual preprocessing. Table~\ref{tab:exp6-scores} reports the measured task scores on a 0--100 scale; VisualWebBench is the equal-weight mean of its seven subsets.

\begin{table}[!htbp]
  \caption{Task accuracy under the deployment quantization configurations (0--100). Each cell lists a/b/c: a = per-channel INT8; b = HQQ INT4 (block size 64); c = symmetric Q4\_0 INT4 (block size 32). Vision weights use INT8 (Q8\_0 for c), except NEO1.5's vision weights are unquantized for c.}
  \label{tab:exp6-scores}
  \centering
  \footnotesize
  \newcommand{\quantcell}[1]{{\fontsize{\dimexpr\csname f@size\endcsname pt-1pt\relax}{\baselineskip}\selectfont\mbox{#1}}}
  \setlength{\tabcolsep}{2pt}
  \begin{tabular*}{\linewidth}{@{\extracolsep{\fill}}l*{6}{c}@{}}
    \toprule
    Task & \shortstack{Qwen3.5-\\2B} & \shortstack{Gemma4-\\E2B} & \shortstack{InternVL3.5-\\1B} & \shortstack{SmolVLM2-\\500M} & \shortstack{NEO1.5-\\2B-SFT} & \shortstack{Mono-\\InternVL-2B} \\
    \midrule
    MMMU & \quantcell{38.44\slash37.22\slash37.00} & \quantcell{32.44\slash31.44\slash29.00} & \quantcell{39.89\slash39.78\slash37.00} & \quantcell{28.22\slash26.56\slash28.78} & \quantcell{49.67\slash47.89\slash48.00} & \quantcell{31.33\slash32.11\slash30.00} \\
    MMStar & \quantcell{41.95\slash42.61\slash35.16} & \quantcell{31.00\slash30.97\slash28.41} & \quantcell{49.48\slash48.07\slash42.78} & \quantcell{37.69\slash31.46\slash33.44} & \quantcell{59.55\slash55.56\slash57.78} & \quantcell{45.95\slash42.06\slash42.66} \\
    SQA-Img & \quantcell{82.05\slash78.68\slash62.96} & \quantcell{53.59\slash57.86\slash43.33} & \quantcell{85.28\slash77.49\slash80.07} & \quantcell{73.77\slash64.01\slash56.72} & \quantcell{96.23\slash95.29\slash95.49} & \quantcell{93.41\slash93.16\slash92.81} \\
    RealWorldQA & \quantcell{67.19\slash66.27\slash61.70} & \quantcell{45.49\slash46.67\slash49.41} & \quantcell{57.52\slash52.68\slash54.64} & \quantcell{51.63\slash48.50\slash44.44} & \quantcell{62.48\slash60.52\slash62.22} & \quantcell{55.69\slash51.37\slash51.37} \\
    OCRBench & \quantcell{82.10\slash76.60\slash79.50} & \quantcell{66.60\slash62.40\slash63.50} & \quantcell{75.30\slash70.10\slash73.30} & \quantcell{60.60\slash53.60\slash54.10} & \quantcell{77.10\slash76.30\slash74.60} & \quantcell{73.90\slash68.00\slash67.80} \\
    MathKangaroo & \quantcell{21.14\slash19.48\slash12.68} & \quantcell{13.58\slash14.73\slash12.90} & \quantcell{21.09\slash12.48\slash18.44} & \quantcell{18.58\slash18.16\slash16.73} & \quantcell{21.34\slash21.17\slash21.65} & \quantcell{17.82\slash18.27\slash17.15} \\
    MTVQA & \quantcell{27.31\slash26.81\slash26.72} & \quantcell{21.26\slash19.59\slash17.10} & \quantcell{19.65\slash17.67\slash17.79} & \quantcell{7.66\slash5.98\slash5.65} & \quantcell{21.51\slash20.58\slash20.55} & \quantcell{8.97\slash7.54\slash6.88} \\
    VisualWebBench & \quantcell{47.59\slash45.34\slash48.93} & \quantcell{34.25\slash32.08\slash34.34} & \quantcell{35.29\slash28.50\slash31.60} & \quantcell{21.91\slash13.51\slash17.48} & \quantcell{44.13\slash40.80\slash40.13} & \quantcell{23.28\slash21.36\slash17.40} \\
    ViewSpatial & \quantcell{39.02\slash39.48\slash39.13} & \quantcell{28.82\slash29.74\slash26.93} & \quantcell{33.46\slash25.04\slash30.43} & \quantcell{23.60\slash23.34\slash23.98} & \quantcell{52.92\slash50.70\slash52.71} & \quantcell{34.38\slash32.95\slash30.50} \\
    MME-RealWorld & \quantcell{31.29\slash42.31\slash43.25} & \quantcell{40.72\slash36.99\slash36.77} & \quantcell{40.98\slash36.35\slash36.75} & \quantcell{35.72\slash29.36\slash27.58} & \quantcell{55.98\slash51.67\slash54.13} & \quantcell{41.48\slash33.23\slash37.30} \\
    \bottomrule
  \end{tabular*}
  \par\smallskip
  {\scriptsize\raggedright Failed MMMU samples are scored as zero; ViewSpatial samples exceeding the context limit are excluded, so the evaluated sample sets can differ between configurations.\par}
\end{table}

\end{document}